\documentclass[letterpaper,twocolumn,10pt]{article}
\usepackage{usenix2019_v3}

\usepackage{tikz}
\usepackage{amsmath}
\usepackage{tikz}
\usepackage{multirow}
\usepackage{amsmath}
\usepackage{amssymb}
\usepackage{placeins}
\usepackage{nicefrac}
\usepackage{xspace}
\usepackage{caption}
\usepackage{numprint}
\usepackage{subcaption}
\usepackage{graphicx}
\usepackage[shortlabels]{enumitem}
\newcommand{\sect}{§}

\newcommand{\etal}{\emph{et~al.}\xspace}
\newcommand{\sota}{state-of-the-art\xspace}
\newcommand{\adversary}{\ensuremath{\mathcal{A}}\xspace}
\newcommand{\challenge}[1]{C#1}
\newcommand{\criteria}[1]{R#1}
\newcommand{\precision}{Pr\xspace}

\newcommand{\recall}{Re\xspace}
\newcommand{\fScore}{F1-Score\xspace}
\newcommand{\fScores}{F1-Scores\xspace}
\newcommand{\fpr}{FPR\xspace}

\newcommand{\devicePresent}{$\bullet$}
\newcommand{\deviceNotPresent}{$\circ$}

\newcommand{\ournameNoSpace}{\mbox{ARGUS}}
\newcommand{\ourname}{\ournameNoSpace\xspace}
\newcommand{\ournameGen}{\ournameNoSpace's\xspace}
\newcommand{\paperTitle}{\ournameNoSpace: Context-Based Detection of Stealthy IoT Infiltration Attacks} 

\newcommand{\hapShort}{HAP\xspace}
\newcommand{\hapLong}{home automation platform\xspace}
\newcommand{\hapF}{\hapLong (\hapShort)\xspace}

\newcommand{\monitoringComponentNoSpace}{Device Monitoring}
\newcommand{\neuralNetworkComponentNoSpace}{Context Modeling}
\newcommand{\classificationComponentNoSpace}{Anomaly Score Classification}
\newcommand{\monitoringComponent}{\monitoringComponentNoSpace\xspace}
\newcommand{\neuralNetworkComponent}{\neuralNetworkComponentNoSpace\xspace}
\newcommand{\classificationComponent}{\classificationComponentNoSpace\xspace}
\newcommand{\windowLength}{\ensuremath{\mathit{l}}}
\newcommand{\datasetHomeOne}{Home~1\xspace}
\newcommand{\datasetHomeThree}{Home~3\xspace}
\newcommand{\datasetHomeTwo}{Home~2\xspace}
\newcommand{\datasetHomeFour}{Home~4\xspace}
\newcommand{\datasetHomeFive}{Home~5\xspace}
\newcommand{\numSetups}{5\xspace}

\definecolor{goodColor}{rgb}{0.196, 0.8156 0.196}
\definecolor{badColor}{rgb}{0.76, 0.094 0.027}
\newcommand{\presentSign}{{\large{\textbf{\checkmark}}}}
\newcommand{\presentGood}{\color{goodColor}{\presentSign}}
\newcommand{\presentBad}{\color{badColor}{\presentSign}}
\newcommand{\absenceSign}{{\large{\textbf{\ensuremath{\pmb{\times}}}}}}
\newcommand{\absentGood}{\color{goodColor}{\absenceSign}}
\newcommand{\specialcell}[2][c]{
  \begin{tabular}[#1]{@{}c@{}}#2\end{tabular}}
\newcommand{\absentBad}{\color{badColor}{\absenceSign}}

\begin{document}

\date{}

\title{\paperTitle}

\author{
\rm Phillip Rieger \\ ~\\\rm Markus Miettinen
\and
\hspace{-0.5cm}\rm Marco Chilese \\ ~\\\hspace{-0.5cm}\rm Hossein Fereidooni\\ ~\\Technical University of Darmstadt
\and
\hspace{-0.5cm}\rm Reham Mohamed \\ ~\\\hspace{-0.5cm}\rm Ahmad-Reza Sadeghi
} 

\maketitle

\begin{abstract}
\noindent IoT application domains, device diversity and connectivity are rapidly growing. IoT devices control various functions in smart homes and buildings, smart cities, and smart factories, making these devices an attractive target for attackers.      

\noindent On the other hand, the large variability of different application scenarios and inherent heterogeneity of devices make it very challenging to reliably detect abnormal IoT device behaviors and distinguish these from benign behaviors. Existing approaches for detecting attacks are mostly limited to attacks directly compromising individual IoT devices, or, require predefined detection policies. They cannot detect attacks that utilize the control plane of the IoT system to trigger actions in an unintended/malicious context, e.g., opening a smart lock while the smart home residents are absent.

\noindent In this paper, we tackle this problem and propose \ourname, the first self-learning intrusion detection system for detecting \emph{contextual attacks} on IoT environments, in which the attacker maliciously invokes IoT device actions to reach its goals. 

\noindent \ourname monitors the contextual setting based on the state and actions of IoT devices in the environment. An unsupervised Deep Neural Network (DNN) is used for modeling the typical contextual device behavior and detecting actions taking place in abnormal contextual settings. This unsupervised approach ensures that \ourname is not restricted to detecting previously known attacks but is also able to detect new attacks. We evaluated \ourname on heterogeneous real-world smart-home settings and achieve at least an F1-Score of 99.64\% for each setup, with a false positive rate (\fpr) of at most 0.03\%. 
\end{abstract}

\section{Introduction}
\label{sect:intro}
IoT devices are becoming an integral part of modern life in many application domains like smart homes, smart buildings, smart city infrastructure, and smart factories. Increasingly IoT devices are also providing access to rich contextual information, making it possible to realize intelligent ambient environments in which whole systems of interconnected IoT devices are controlled in a coordinated and adaptive way. For instance, IoT devices can sense data about the movements and behavior of smart home users and automatically adapt lighting, heating, or air conditioning settings accordingly.

In 2020, more than 11.3 billion IoT devices were deployed in smart homes and more than 27 billion devices are expected for 2025~\cite{iotstats}. In the US, 23\% of all broadband households have already 3 or more connected devices~\cite{iotstats3} and one can expect this number to increase since more and more functions will be controlled by IoT devices.

\noindent \textbf{Attacks on IoT Systems.} The continuously growing number and diversity of IoT devices from different device manufacturers enlarges the potential attack surface in many IoT networks. While different approaches have been proposed to detect attacks that directly compromise IoT devices, e.g., through IoT malware~\cite{diot,raza2013svelte,JiaNDSS2017ContexIOT}, more stealthy attacks compromising the \emph{contextual integrity}~\cite{contextualIntegrity2006} of IoT networks by misusing the \emph{control plane} of IoT devices, i.e., local and remote systems and applications like vendor-provided smartphone apps or cloud services that are used to control the IoT devices, have not been sufficiently addressed yet. Detecting attacks on contextual integrity is highly challenging. Since both, benign actions and \emph{contextual attacks}, use regular control commands for triggering legitimate actions and also the network traffic does not necessarily differ, benign and attack-actions are as such indistinguishable. The only difference between them is the context of their invocations, identified through the environmental factors and stages of the other devices in the smart-home. Therefore the attack-actions are invoked in a situation, where the user does not desire it.

For instance, an attacker who infiltrated the vendor's cloud service, e.g., due to weak credentials, could instruct a smart lock in the apartment of the victim user to open the door while the user is absent so that the attacker can then break into the user's home. The attacker could also compromise the safety of a smart home by sending a command to turn on the smart stove, causing a potential fire hazard. Also other, non-intentional device or system failures may compromise the safety of the target environment, e.g., if the IoT-controlled heating system fails and turns off completely while the user is on vacation, such that the indoor temperature falls below the freezing point. 
Unlocking the door, turning off the heating, or, turning on the stove, considered individually, are perfectly normal actions. Consequently, if the attacker uses the control plane to trigger these actions, the related commands \mbox{look like} perfectly benign commands. Since the commands are sent in these examples by the vendor-specific cloud platform, the network traffic does not differ from the traffic of benign actions. Therefore, in order to detect such attacks, also the \emph{context} of actions that IoT devices undertake needs to be considered.

\noindent\textbf{Existing Defenses.} Existing approaches for IoT intrusion detection systems often focus on analyzing the network traffic~\cite{homesnitch,diot,fan2020iotdefender}. However, these approaches cannot detect contextual attacks, as the network traffic of attack commands is indistinguishable from benign commands. The remaining approaches that consider contextual information fall into different categories. The most relevant categories are: 1) Validation of Sensor Values, 2) Policy Enforcement (either defined rules or dynamically through graph representations), and 3) systems for Contextual Anomaly Detection. 
The systems in the first category~\cite{adkisson2021autoencoder,yasaei2020iot,yin2020anomaly,kotevska2019kensor} often focus on detecting anomalies only with regard to a specific sensor. This enables them to detect wrong measured sensor values, but making them fail to recognize abnormal physical system states. For example, if the door of an apartment is physically opened during the users' absence, then the values of the sensors are correct when indicating this, although opening the door while nobody is at home should be considered anomalous.

Approaches in the second category use fixed policies ~\cite{yamauchi2020anomaly,feng2019systematic,yahyazadeh2019expat,celik2018soteria,choi2018detecting,kafle2021towards} that are often determined by non-dynamic processes. A user must either extract these policies from the source code/description of apps that control IoT devices, ignoring the actual user behavior, or they must be manually defined by the users, which is inconvenient.

Existing approaches that fall into the third category by training anomaly detection modules based on captured training data~\cite{fu2021hawatcher,amraoui2020ml,tang2019smart,sikder2019aegis,sikder20176thsense} also suffer from the requirement of semantic information about the IoT devices~\cite{fu2021hawatcher}, being restricted to having example data for the attacks~\cite{tang2019smart}, to analyzing only the commands~\cite{amraoui2020ml}, or cannot model the relationships between the individual devices accurately~\cite{sikder20176thsense,sikder2019aegis}.

An autonomous approach that monitors the context of the IoT devices without requiring access to source code is therefore needed. If the attack is detected in time, the user can take countermeasures in time to mitigate the attack, e.g., by calling the police if the door lock is opened while she is absent.

\noindent\textbf{Our Approach.} In this paper, we propose \ourname, a novel approach for detecting contextual attacks against IoT networks, i.e., attacks which perform benign actions in a wrong context. \ourname is a complementary solution augmenting existing network traffic monitoring-based intrusion detection approaches that focus on detecting direct attacks on IoT devices (e.g., IoT malware attacks) but cannot detect contextual attacks utilizing the control plane. 
To detect such stealthy, contextual attacks, \ourname is inspired by the notion of \emph{contextual integrity}~\cite{contextualIntegrity2006} and makes use of the observation that the acceptability and permissibility of an action are highly dependent on the contextual setting in which the action is taking place. 
It models the context in terms of events, user actions, device actions, and triggered automation rules to overcome the limitation of network-traffic-based approaches. Another challenge that \ourname addresses is that what is considered "normal" behavior is highly dependent on individual networks and users, preventing the use of simple static policies for distinguishing between malicious and benign actions. 
To address this challenge, \ourname trains a Deep Neural Network (DNN) for capturing the interdependence between contextual factors and events and device actions. 
With the help of the trained DNN, \ourname evaluates the actions of IoT devices in the network and determines for each action an anomaly score to detect anomalous situations in which an invoked action is not consistent with the \emph{contextual setting} in which it is occurring.
Our contributions are as follows:
\begin{itemize}
\itemsep0em 
    \item We present \ourname, a context-based intrusion detection framework for IoT networks capable of detecting IoT infiltration attacks in which the adversary compromises the control plane of the network, e.g., cloud servers or mobile apps, to stealthily manipulate the behavior of devices to achieve malicious goals. Thus, \ourname detects contextual attacks where the individual device actions are normal but are performed in a wrong context~(\sect\ref{sect:design-highLevel}).
    \item We develop a dynamic tuning scheme for the classification boundary of events' anomaly scores that automatically adapts to different setups (\sect\ref{sect:approach-threshold}).
    \item We collect and provide the first real-world dataset, capturing the behavior of different smart-homes to be utilized by research community for conducting future research in this area\footnote{\url{https://github.com/TRUST-TUDa/argus-data}} (\sect\ref{sect:eval-datasets}).
    \item We extensively evaluate \ourname on the collected real-world dataset, consisting of \numSetups heterogeneous smart-home setups~(\sect\ref{sect:eval-results}).
\end{itemize}

\section{Problem Setting}
\label{sect:problem}
In this section, we first briefly elaborate on recent IoT attacks and then explain our system model and design to mitigate such attacks. Afterward, we describe the contextual threat model and challenges that \ourname is designed to solve.

\begin{figure}[t]
    \centering
    \includegraphics[width=0.7\columnwidth]{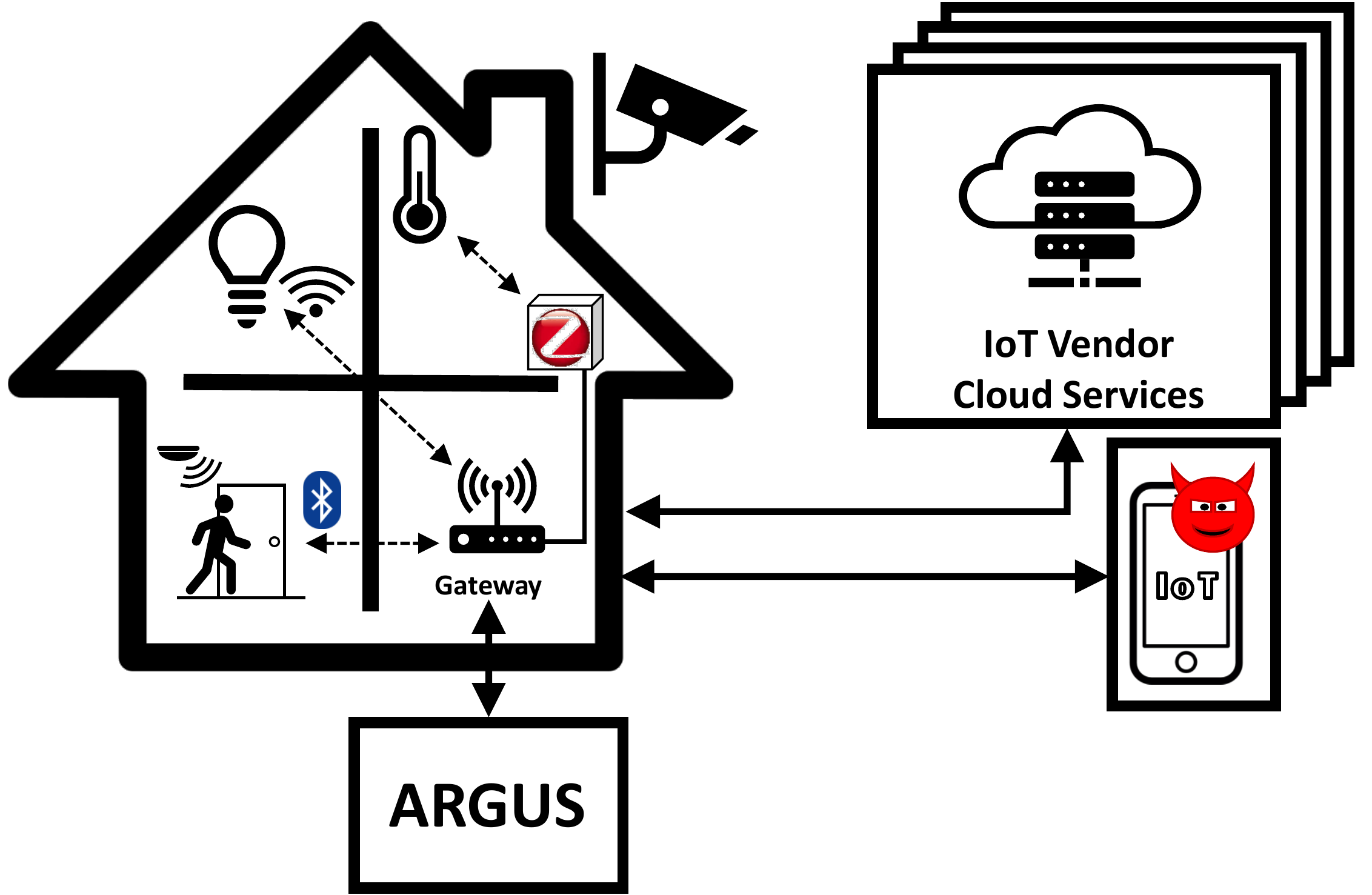}
    \caption{\ourname system model}
    \label{fig:systemmodel}
\end{figure}
\subsection{Recent IoT Attacks}

In the course of IoT market proliferation, an increasing number of attacks on IoT devices have been reported~\cite{o2016insecurity}. Some prominent attacks were related to Mirai botnet~\cite{antonakakis2017usenixMirai} and its successors~\cite{herwig2019measurement}, where a massive number of compromised IoT devices was used to stage one of the largest DDoS attacks ever recorded on the internet. However, recently other attacks have emerged that specifically target functions controlled by IoT devices, for instance, a botnet of IoT devices was able to effectively incapacitate the heating system of a large residential building while the temperature outside was far below freezing~\cite{finland}. Obviously, such attacks could have even more damaging consequences, if instead of a heating system, e.g., vital devices in a hospital would be targeted. 

These examples demonstrate the need for effective countermeasures against attacks targeting the \emph{control plane} of IoT networks, i.e., the systems, protocols, and mechanisms controlling the functionality of the IoT devices.

\subsection{System and Context Model}
\label{sect:problem-system}
Our system model, shown in Fig.~\ref{fig:systemmodel}, considers a heterogeneous IoT network consisting of different IoT devices controlling various functions related to their ambient environment and measuring different parameters of the environment and their operational context. Some of the IoT devices may be associated with vendor-specific IoT cloud services and may utilize associated mobile applications for allowing the user to remotely control these devices.
All devices are connected to the local network, which typically also provides access to the internet via an access gateway. Devices connect to the local network using Ethernet or WiFi, or, a specific hub device providing IP-connectivity for devices using wireless proximity protocols like Bluetooth, ZigBee, or Z-Wave.

Via the internet, the IoT devices are connected to the control plane, consisting of all local and remote systems and applications that are used to legitimately control the IoT devices.  In the rest of the paper, we assume that the attacker compromises one of these entities (cf.~\sect\ref{sect:problem-adversary}).

An example of such a setup is shown in Fig.~\ref{fig:systemmodel}. Here, the smart-home consists of a light-bulb (connected through WiFi), a smart-lock (connected through Bluetooth), an IP-Camera (connected through LAN), and a smart thermometer (connected through ZigBee via a hub). Each device uses a different cloud platform and the user controls them via vendor-specific applications on its mobile phone. The cloud platforms and the mobile applications represent here the control plane of the setup.
The goal of \ourname is to monitor the actions taken by individual IoT devices, e.g., turning on the light or unlocking the door, and notify the user in case it detects device actions that are not consistent with the current contextual setting.

The main focus of \ourname are \emph{contextual attacks} where the attacker uses benign functions of IoT devices triggered in incorrect contextual settings to stage attacks for achieving its attack goals.
For detecting such attacks, we model the context of the IoT network in terms of a number of context features characterizing the system's contextual state. The considered contextual features can be roughly classified into three main categories: 1) Ambient and temporal features describing the environment (noise level, luminosity, humidity, temperature, time of the day, etc.), 2) Features indicating the context of the user (asleep/awake, present/absent,  etc.), 3) Device states (device state changes, triggered automation rules, event notifications, device alarms, etc.).

\noindent The context features can be automatically harvested from the monitored IoT system using appropriate APIs of individual IoT devices, their associated cloud services, and possible home automation systems installed in the user's network. 
As described in more detail in \sect\ref{sect:design}, \ourname aggregates these contextual factors and feeds them to a machine learning algorithm used to profile the contextual state of the local IoT setup and perform anomaly detection.

\subsection{Adversary Model}\label{sect:problem-adversary}

\begin{figure*}[t]
    \centering
    \includegraphics[width=0.65\textwidth]{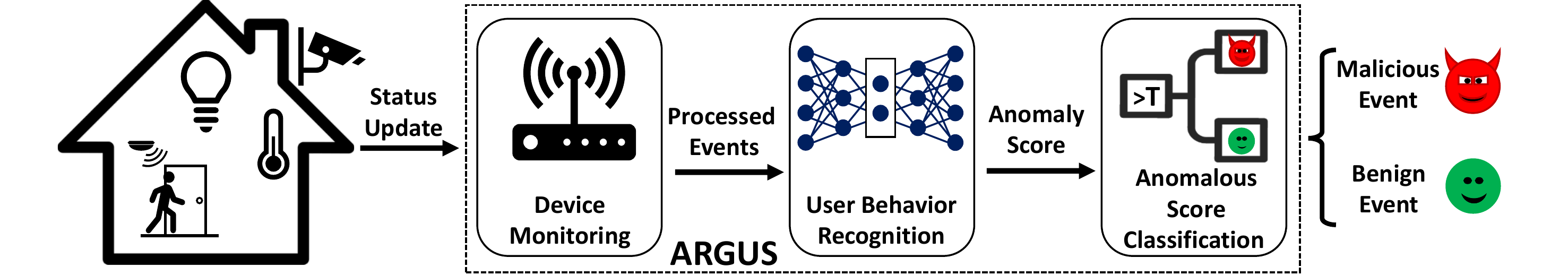}
    \caption{High-Level Overview of the components of \ourname}
    \label{fig:highlevelSystemOverview}
\end{figure*}

\noindent We consider an adversary \adversary that compromises a part of the IoT control plane to trigger normal-looking actions on the IoT devices but in the wrong context. Therefore, the legitimate user does not want these actions to be executed. Note that for staging the attacks \adversary does not need to actually compromise the targeted IoT devices and execute malicious code locally on the IoT device itself. It is sufficient to abuse the compromised control plane to invoke commands. The invoked commands look, when considering them individually, benign and might, in another context, be also invoked by the user itself. Therefore, only the context of invocation allows to distinguish benign actions and attacks.

For compromising the control plane, there exist a number of attack vectors, e.g., when the IoT device or the related cloud service use insufficient authentication (weak, default~\cite{pentestHotTubeSamePassword}, or even missing passwords~\cite{pentestCameraNoPassword}), or a malware placed on the smartphone of the user targeting specific IoT apps.

An example of a contextual attack is shown in Fig.~\ref{fig:systemmodel}. \adversary compromises an app on the user's mobile phone, which is part of the control plane and uses it to post a control command to the smart lock, causing it to open the door while the user is asleep or not at home. In general, unlocking the door using the mobile phone is a benign event, which is also invoked by the user. However, the normal context for such actions would be, e.g., that the lock opens when the user is returning home and approaching the door. The fact that the user is asleep, or, absent from home clearly represents an abnormal context for the smart lock action of opening the door.

The motivation of \adversary for the attack is to invade the users' privacy, cause financial damages to the user, or harm the user in some other way.

However, we assume that \adversary neither compromises the actual IoT devices nor \ourname, as we will elaborate in our trust model (cf.~\sect\ref{sect:design-trust}).

\subsection{Requirements and Challenges}
\label{sect:problem-requirements}

For detecting contextual attacks in realistic real-world settings in an effective and user-friendly manner, the \ourname system should satisfy the following requirements:\\
\textbf{\criteria{1} Fast detection:} Since contextual attacks can potentially lead to significant physical damages or monetary loss to the user, attacks must be detected in near real-time when new incoming events take place in the system in order to allow sufficient reaction time for taking appropriate countermeasures.\\
\textbf{\criteria{2} Cause identification:} The system must be able to identify the device or event that is causing an alarm to be triggered. 
This is necessary to allow the user to understand what the root cause is and choose appropriate countermeasures to mitigate the attack.\\
\textbf{\criteria{3} Minimizing false alarms:} The system must not generate many false alarms to ensure that the user is not overwhelmed with false alarm notifications. Otherwise, the user will likely start ignoring incoming alerts, or, disable the protection system altogether to avoid unnecessary inconveniences.\\ 
\textbf{\criteria{4} Autonomous operation:} The system must run with minimal configuration input from the user. Users should be required to perform only the basic configurations while the system should take care of training and applying the necessary contextual models for attack detection. Otherwise, the system will not be practical as users are not likely to have sufficient expertise and be willing to spend considerable effort in configuring the system.\\

\noindent To cater to the above requirements, the system must thus solve the following technical challenges:\\
\noindent\textbf{\challenge{1} Detecting attacks consisting of benign actions:} As discussed in \sect\ref{sect:problem-system}, contextual attacks constitute situations, in which the adversary triggers actions in wrong contextual situations where the actions themselves represent legitimate benign actions of the device. The main challenge is how one can detect an attack consisting of \emph{benign} actions? Some existing approaches monitor the network traffic of IoT devices to detect known attack patterns, or, deviating traffic patterns caused by potential attacks~\cite{diot,fan2020iotdefender,homesnitch,iotgaze,raza2013svelte,JiaNDSS2017ContexIOT}. In the setting of contextual attacks, these approaches are, however, not applicable, as the traffic patterns used to stage the attack represent in essence 'benign' operations of the devices.

\noindent\ourname seeks to resolve this challenge by utilizing a context model detailed in \sect\ref{sect:approach} that explicitly links device actions with the contextual setting they take place in. This allows the detection model to enforce contextual integrity by learning what sequences of events are benign, normal event sequences, and what represent potential attacks. This also allows to detect attacks immediately when they occur (\criteria{1}) and to specifically indicate, which particular action it was that triggered detection, thus helping in identifying what device action caused a potential alarm (\criteria{2}).

\noindent\textbf{\challenge{2} Autonomous defense personalization:} For minimizing the number of false  alarms (\criteria{3}), the detection system must be tailored to the local IoT set-up and personalized to consider personal preferences and habits of users. The challenge here is, how this can be done without requiring extensive manual configuration of enforcement policies to enable autonomous operation of the system (\criteria{4})?
Earlier systems addressing contextual attacks heavily depend on explicit input from the user to define policies determining permissible and undesired IoT device actions in particular contextual situations~\cite{iotgaze,celik2019iotguard,tian2017smartauth,iotsafe}. 
Such approaches are unlikely to work in general in real-life settings, since regular IoT users are very unlikely to have the required experience or motivation to spend a considerable amount of time and effort in setting up secure and effective security policies for their IoT devices that sufficiently accurately match their IoT set-up and personal preferences.
\ourname seeks to tackle this challenge by replacing pre-specified static policies by a trained detection model that can be trained requiring only minimal explicit inputs from users.

\section{System Design}
\label{sect:design}
In the following, we provide a high-level overview of \ourname, the first context-based IoT intrusion detection framework that monitors connected IoT devices together with their context to detect abuse. \ourname uses an anomaly detection approach that compares newly captured events against the modeled behavior to detect anomalous actions (e.g., unseen attacks). In \sect\ref{sect:design-trust} we elaborate on the considered trust model and security assumptions.
\subsection{High-Level Overview}
\label{sect:design-highLevel}

A high-level overview of \ourname is depicted in Fig.~\ref{fig:highlevelSystemOverview}. It involves the following entities: \monitoringComponent, \neuralNetworkComponent, and \classificationComponent components to monitor the actions of the IoT devices and detect anomalous behavior. In the following, we outline the role of each component. Details for each component are discussed in \sect\ref{sect:approach}.\\
\textbf{\monitoringComponent.} The monitoring component collects the status updates and event notifications from the observed IoT devices and preprocesses them for the following components. For managing the different APIs of different IoT ecosystems, \ourname makes use of a home automation platform that connects to the individual APIs of the various IoT devices installed in the system.\\
\textbf{\neuralNetworkComponent.} This component utilizes an Auto-Encoder~(AE) architecture~\cite{bourlard1988auto} which is an unsupervised DNN approach, since it uses only benign data for training. The AE predicts anomaly scores for each new incoming event that is captured by the \monitoringComponent component. The score indicates the similarity of the captured events with the modeled expected behavior.\\
\textbf{\classificationComponent.} The final component of \ourname uses a dynamically calculated threshold as classification boundary to discriminate each event based on its anomaly score as benign or malicious. The threshold is based on the anomaly scores of the previous day as well as the previous threshold. An event is considered to be an attack, if the anomaly score is higher than the threshold.
\subsection{Security and Trust Assumptions}
\label{sect:design-trust}
We make following assumptions with regard to the capabilities of the adversary \adversary and the trusted system components:

\begin{itemize}
\itemsep0em 
    \item Aligned with existing work \ourname considers only attacks that compromise the control plane of the IoT network but not direct attacks against the IoT devices themselves~\cite{kafle2021towards}. This is because there exists a large body of work focusing on detecting direct attacks against IoT devices~\cite{diot,fan2020iotdefender,homesnitch,iotgaze,raza2013svelte,JiaNDSS2017ContexIOT}. \ourname should thus be seen as a complementary approach augmenting these defenses by providing the ability to detect also contextual attacks targeting the control plane. Consequently, since \adversary cannot compromise the IoT devices, e.g., to run code locally, it also cannot suppress or fake status updates. It also cannot utilize IoT devices to impersonate other IoT devices~\cite{golomb2018ciota,sikder2019aegis}, as this would require compromising the utilized IoT devices first.
    \item Aligned with previous work~\cite{celik2019iotguard,yahyazadeh2019expat}, we assume the individual components of \ourname to be trusted. While during the development of IoT devices there is less focus on security considerations, \ourname is specifically designed to increase the security, such that security experts are involved in its implementation and also it is reasonable to expect the system to be hardened for security. Because of the focus on security considerations during the implementation and the limited set of functions, we assume that \adversary \mbox{cannot compromise the components of \ourname.}
    \item Aligned with previous work~\cite{diot,fu2021hawatcher}, we assume the local IoT setup to be not compromised during training time.
\end{itemize}

\section{\ourname}
\label{sect:approach}
In the following we describe the individual components of \ourname (\monitoringComponent, \neuralNetworkComponent, \classificationComponent) in detail.
\subsection{\monitoringComponent}
\label{sect:approach-monitoring}
The \monitoringComponent component collects status updates from the individual IoT devices, allowing \ourname to also consider the context of an invoked action to determine whether the action is benign (addressing~\challenge{1}). A challenge here is the very heterogeneous IoT device landscape. Different devices from different manufacturers use different protocols, are connected to different cloud platforms and mobile applications, and might even use different communication technologies. For example, one device might use WiFi to connect to the local network while other devices might use ZigBee and must use a dedicated ZigBee hub device as a gateway to connect to the local network. For these reasons, most existing intrusion detection systems (IDS) consider only the LAN and WiFi traffic but do not consider other wireless communication protocols~\cite{diot,homesnitch,fan2020iotdefender}.

\begin{table}[b]
    \caption{Model Hyperparameters}
    \label{tab:hyperparam}
    \centering
    \begin{tabular}{l|cl}
        Variable & Setting \\ \hline
        Encoder Layer Type  & [GRU, GRU] \\ 
        Decoder Layer Type  & [GRU, GRU] \\ 
        Encoder Hidden Units& [GRU:256, GRU:64]  \\
        Decoder Hidden Units& [GRU:64, GRU:256]  \\ 
        Optimizer &  Adam \\ 
        Loss Function & MSE \\ 
        Learning Rate & Decaying from $10^{-3}$ to $10^{-6}$ \\
        Dropout Value  &  0.3 \\ 
        Epochs & Max \numprint{35000}\\ 
        Batch Size & 64 \\ 
    \end{tabular}
\end{table}
The \monitoringComponent component of \ourname addresses this challenge by making use of a home automation system. A number of such systems that integrate various IoT devices and allow to control them in one central place have been developed in the past~\cite{homeassistant, appleHome,samsungThings,googleHome}. While some manufacturers might put less focus on security aspects, an easy integration and access is vital for deployment and user acceptance of IoT devices creating a strong motivation for manufactuers to allow a simple integration into existing HAPs.
The \monitoringComponent of \ourname exploits this by utilizing the automation system for the integration of the devices, maintaining the adapters to use the APIs of the different IoT ecosystems, and collecting the events from IoT devices of different vendors and technologies. This allows \ourname to use all available devices for monitoring the context.

When a device changes its status, the device reports this status update to the home automation system. The remaining part of the \monitoringComponent component then records the status update, parses the status if necessary to ensure a standard data format, adds the new event to the sequence of previous events, orders all events by their occurrence time, and forwards them to the \neuralNetworkComponent component.

\subsection{\neuralNetworkComponent}
\noindent In order to distinguish between a normal scenario and a suspicious event in view of the usual user and system behavior, the \neuralNetworkComponent component of \ourname models the expected behavior and its context based on the previously collected training data.
\noindent In the following subsections, we will describe the data preprocessing as well as the architecture of the Deep Neural Network (DNN) that is used for modeling the user's expected behavior.

\subsubsection{Data Preprocessing}
\label{sect:approach-ml-preprocessing}
The data preprocessing is performed in the following steps:
\begin{enumerate}
	\item Parsing log data: The captured events are parsed in order to store the information about the devices' and sensors' status update events. For the ease of presentation, in the following device states refer to the states of the individual IoT devices as well as the captured values of the individual sensors.

	\item States' value mapping: In order to deal with the multitude of devices' states and for facilitating the subsequent phase of ML, all device states are mapped into numeric values in a restricted range. Devices that have only a limited or nominal set of states (e.g., "on" and "off") are mapped in range $[0,1]$. Each observed state is first mapped to a cardinal number $\text{state}_{id}$, which is then normalized corresponding to:
	\begin{equation}
	    S_i=\frac1{|\text{states}|}\cdot\text{state}_{i_{\text{id}}}
	\end{equation}
    where $S_i$ is the state $i$ in the set of states for that device, $|\text{states}|$ is the set of all states, and $\text{state}_{i_{id}}$ is the cardinal number of that state in the set. The state $S_0$ is reserved for new values that were not observed during the training phase. Continuous values (e.g., temperature and humidity values) are mapped to 10 values in range $[0,1]$ corresponding to $[\text{state}_{min},\text{state}_{max}]$.
    The interval $r_i$ for the values that are mapped to value $\nicefrac{i}{10}$ is given by :
	\begin{equation}
	    r_i =\left[S_{\min}+i\frac{S_{\max}-S_{\min}}{10}, \;S_{\min}+\frac{S_{\max}-S_{\min}}{10}(i+1)\right]
	\end{equation}
	where $S_{\max|\min}$ is the max or min value of the states' set of that device, and $i \in \{0,\ldots, 9\}$ is the numerical value of each bucket of the new mapping, reserving $S_0$ for future unseen values.
	
	\item Event chain construction: For each moment in time of the recorded events, the state of each device in the system is reconstructed from the status updates, for having a complete view of the system at every moment in time.\\ 
	The resulting chain is characterized by a list of events: $[\text{event}_0, \text{event}_1, \ldots, \text{event}_n]$ where 
	\begin{equation}
	    \text{event}_i = [\text{state}_{\text{device}_0}, \text{state}_{\text{device}_1}, \ldots, \text{state}_{\text{device}_m}]
	\end{equation} in a particular moment in time $i$, where $\text{event}_0$ is the first event recorded and $\text{event}_n$ is the last one.
	
	\item Sequence building: The event chain is converted into event windows of size \windowLength. Therefore, the chain is split accordingly in groups of size \windowLength\xspace for producing feature vectors with shape $(\windowLength, N_{devices})$.
\end{enumerate}

\subsubsection{Deep Learning Model}
\label{sect:approach-ml-model}

\ourname models the users' normal behaviors and the context to distinguish between abnormal and normal behaviors by using an Auto-Encoders~(AE) architecture~\cite{bourlard1988auto} for the Deep Neural Network~(DNN). AEs are widely used in anomaly detection tasks \cite{chen2018autoencoder,zhou2017anomaly,kieu2019outlier}. They consist of two parts, an Encoder and a Decoder. We are using an under-complete AE, so the size of information decreases layer by layer in the Encoder until a "bottleneck" where information reaches the point where the model has extracted all the hidden patterns in data (information at this stage is known as "encoded data"). From this point on, the information will be reconstructed by the Decoder, expanding it layer by layer, to reproduce the input data. Finally, the amount of error made in the reconstruction is measured by using the mean squared error (MSE). The reconstruction \mbox{error of the AE is used as the anomaly score of an event.}

Since the kind of data we are dealing with has a temporal structure we have designed an undercomplete AE made of recurrent unit layers, in particular, based on Gated Recurrent Units (GRU)~\cite{cho2014learning}. The choice of using GRU layers is guided by the fact that we need to be able to learn latent patterns in temporal context for being able to recognize user's behavior in benign scenarios. This enables \ourname to learn even desired random behavior, e.g., randomly turning on/off lights during specific times.

\begin{figure}[t]
    \centering
    \includegraphics[width=.9\columnwidth]{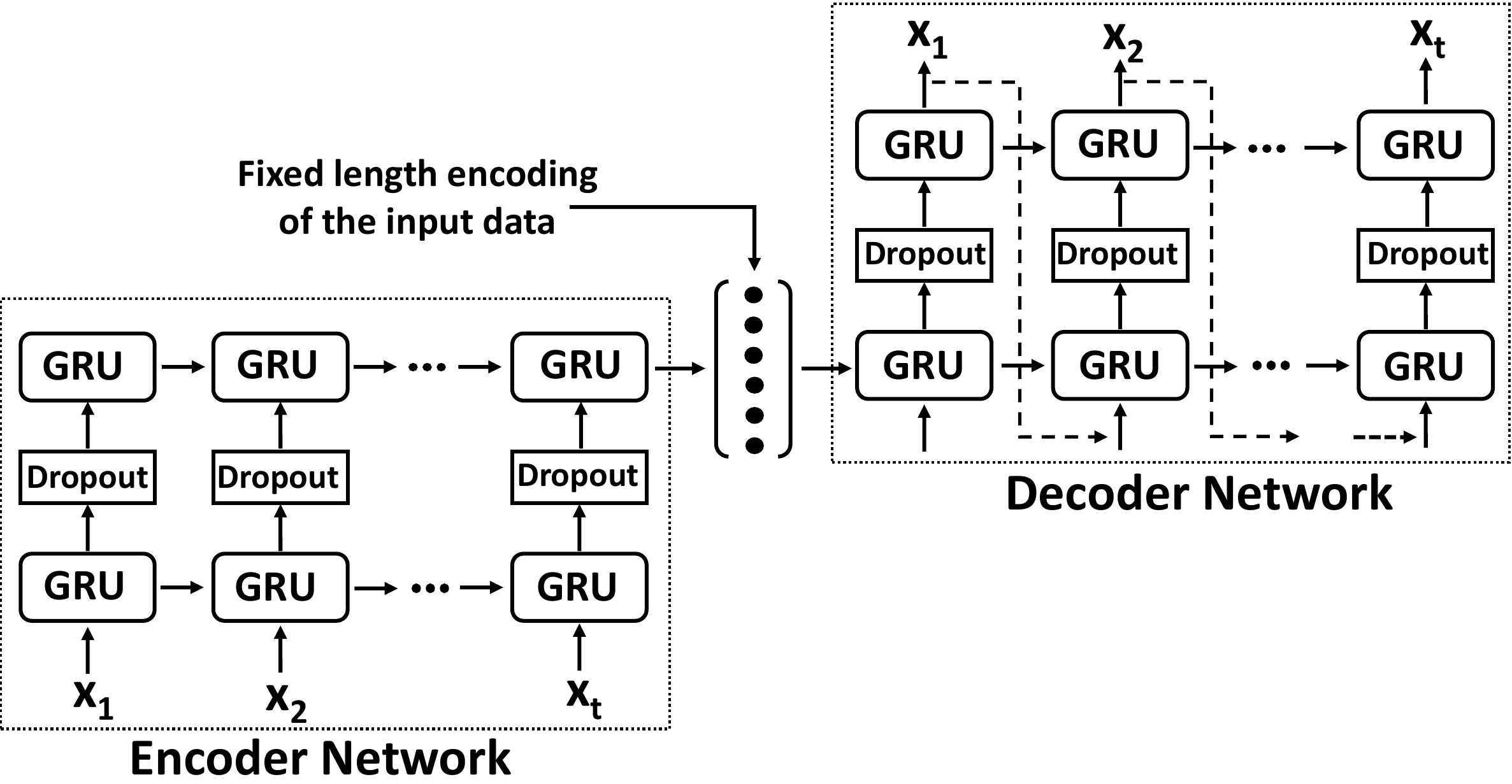}
    \caption{\ourname Auto-Encoder (AE) architecture}
    \label{fig:argusAEOverview}
\end{figure}

Encoder and Decoder are made of two recurrent layers each one, respectively with decreasing and increasing number of hidden units. The architecture of our AE model is depicted in Fig.~\ref{fig:argusAEOverview}. The hyperparameters are shown in Tab.~\ref{tab:hyperparam}.

\noindent The training task concerns reconstructing input data. To do so, the model learns the latent patterns and hidden representation of data, i.e., learning the user's behavior. The AE architecture allows to use only benign data for training the DNN such that the model at end of the process will only be able to reliably reconstruct the benign data (i.e., benign scenarios), while abnormal data, i.e., suspicious events,  cannot be encoded effectively (resulting in a larger reconstruction error) since they are not exposed to the model during the training. A large reconstruction error that exceeds a threshold allows the \ourname system to recognize an event as malicious.

\subsection{\classificationComponent}
\label{sect:approach-threshold}
The \classificationComponent uses the anomaly scores, i.e., the reconstruction errors, which were predicted by the \neuralNetworkComponent component to determine whether the given status update is benign or malicious. To discriminate these values, a threshold $T_d$ is determined dynamically based on the previously observed anomaly scores and used as classification boundary. An event is classified as benign iff its reconstructions error is smaller or equal to $T_d$, otherwise an alarm is raised.\\

\noindent For calculating the threshold $T_d$ that is used on day d + 1, for each previous day $d^* \leq d$ a so-called threshold candidate $C_{d^*}$ is calculated:
\begin{equation}
        C_{d^*} = \max\left(E_{d^*}\right)+\beta\cdot\left(\max\left(E_{d^*}\right) - \min\left(E_{d^*}\right)\right)
        \label{eq:thresholdCandidate}
\end{equation}  
where $E_{d{^*}}$ is the set of reconstruction errors of all events that were collected on day $d^*$ and $\beta$ represents the security level. To prevent that exceptional high reconstruction errors significantly affect the threshold, a momentum \cite{dlbook92542519} is used to combine $C_d$ with the previous threshold $T_{d-1}$. The threshold $T_d$ is then given by:
\begin{equation}
    \begin{array}{l}
        T_d = \begin{cases}
            \alpha \cdot T_{d-1} +(1-\alpha)\cdot C_d & d > 0 \\
            C_0 & d = 0 
        \end{cases}
    \end{array}
\end{equation}  
where $\alpha$ is the aging factor that determines the impact of the previous threshold. The higher $\alpha$ is, the more impact the previous threshold has.\\
Since \ourname captures the events and calculates the anomaly score without any delay, \ourname fulfills by design \criteria{1} (Fast detection). As an anomaly score is determined for every new event, it is possible to identify the reason of the anomaly by presenting the last event as well as a short list of preceding events to the user such that \ourname also fulfills \criteria{2}.

\subsection{Implementation}
\label{sect:approach-implementation}
In the following, we describe the implementation details of the evaluation of \ourname in~\sect\ref{sect:eval}. Further, we define the values for the parameters of \ourname, such that the only adaption that users need to make is the decision, which devices shall be monitored. Therefore, \ourname fulfills \criteria{4} (Autonomous operation).\\
\textbf{\monitoringComponent.} For the monitoring, \ourname utilizes a \hapF that connects different devices and collects their states. In addition, the user can specify and create specific automation rules or configurations for his own network. Using a \hapShort enables \ourname to monitor a large variety of devices from different manufacturers and ecosystems for observing the context. The HAP keeps track of the status of each connected device or sensor in the local home network. Thus, it enables \ourname to monitor the contextual features automatically, without requiring it to set any feature manually.

For the experiments, we used Home Assistant, an open-source platform that supports at the time of writing more than 1000 devices and also protocols like MQTT~\cite{homeassistant}. Users can also control the connected devices from outside the network, e.g., via a mobile app.

A further advantage of the remote access to the HAP is that it allows a flexible deployment of \ourname. Due to the separation of individual components, the \neuralNetworkComponent and \classificationComponent can be placed outside the user's home, e.g., on some trusted cloud servers. Then, only the \monitoringComponent needs to be placed inside the user's network, such that this component can be installed, e.g., on a low-performance device. Alternatively, Especially in the case of privacy concerns of the user, the design and implementation of \ourname also allow a completely local deployment, e.g., on a low-performance device like a RaspberryPi (cf. App.\ref{app:runtime}). By this, \ourname also allows a network setup where no data related to the IoT devices leaves the local network.

\paragraph{\neuralNetworkComponent.} As described in \sect\ref{sect:approach-ml-model}, the \ourname AE architecture is generally applicable and depends only on the number of devices involved in the analysis and on the width of the event window size $\windowLength=16$. By this, \ourname addresses \challenge{2}. Encoder and Decoder are made of two layers each with a fixed number of hidden units, respectively, decreasing (from 256 down to 64) and increasing (from 64 up to 256 for compressing and then reconstructing the input.

For analyzing the user's behavior, the temporal context encoded in the event sequences, needs to be considered. So, we need to take into account the temporal structure of data in order to have a model that is able to learn those patterns. For that reason, we use recurrent layers, in particular, Gated Recurrent Unit (GRU) layers. We used GRUs units rather than alternatives such as LSTMs~\cite{hochreiter1997long} due to their capability of faster convergence, requiring less memory (i.e., less trainable parameters) and better dealing with long-term memory problems (e.g., vanishing/exploding gradient).

The size of the final model depends on the number of devices in the system. In our setups, the number of devices varies from 18 to 40, so the model size varies from 1.2 to 2.7 million trainable parameters.

The designed learning process for such models uses a decaying learning rate (from $10^{-3}$ down to $10^{-6}$) once the number of epochs reaches some specific epochs. Furthermore, we make use of early stopping with patience monitoring validation loss in order to prevent overfitting behaviors.

\paragraph{\classificationComponent.} The threshold that is used as classification boundary to discriminate benign events from attacks makes use for two parameters, the aging factor $\alpha$ and the security level $\beta$.

$\alpha$ is used for performing a trade-off between allowing the threshold to dynamically react to changes in the behavior and preventing high changes in the threshold, which could cause incorrect classifications~(cf.~App.~\ref{app:alphabeta}). We therefore \mbox{set~$\alpha=0.8$.}

The security level $\beta$ ensures an additional margin to prevent unusual behavior of the user from causing false alerts. Too high values prevent \ourname from detecting attacks, while too low values cause false alerts. We empirically set $\beta=0.2$~(cf.~App.~\ref{app:alphabeta}).

\section{Evaluation and Discussion}
\label{sect:eval}
In this section, we discuss how different attacks are applied to the network and how our approach is able to detect such attacks. For the evaluation, we evaluate \numSetups real-world IoT setups. The first subsection shows the modeling of the benign data and the distribution of the testing and training data. The next subsection discusses the real network and how the attacks are performed and detected. For the evaluation, we use the well established performance metrics \fScore, Precision, Recall and False-Positive-Rate (cf. App.~\ref{app:metrics}). In App.~\ref{app:computational-setup} we describe the computational setup, in App.~\ref{app:runtime} we evaluate the runtime performance of \ourname.

\subsection{Dataset}
\label{sect:eval-datasets}
For our experiments, we equipped \numSetups different smart-home settings with a variety of IoT devices and sensors that were used by residents on a daily basis. From each setup, we used the first 7 days for training the model and the remaining data for testing. The data for training were split into 90\% of actual training data and 10\% validation data.

\begin{table}[t]
	\centering
	\caption{Performance of \ourname on Real-World Setups, all values in percentage.}
	\label{tab:eval-results-overview}
		\begin{tabular}{l|rrrrr}
			\multicolumn{1}{c|}{Dataset} & \multicolumn{1}{c}{\fpr} & \multicolumn{1}{c}{\precision} & \multicolumn{1}{c}{\recall} & \multicolumn{1}{c}{\fScore} \\\hline
		  \datasetHomeOne & 0.03 & 99.22 & 100.00 & 99.64\\
            \datasetHomeTwo & 0.00 & 100.00 & 100.00 & 100.00\\
            \datasetHomeThree & 0.00 & 100.00 & 100.00 & 100.00\\
            \datasetHomeFour & 0.00 & 100.00 & 100.00 & 100.00\\
            \datasetHomeFive & 0.00 & 100.00 & 100.00 & 100.00
	\end{tabular}
\end{table}

\subsubsection{Dataset Collection}
\label{sect:eval-datasets-real}
\noindent For collecting the dataset, we captured the status updates of IoT devices in \numSetups different smart-home environments (referred to as \datasetHomeOne\xspace - \datasetHomeFive). Each setup consisted of multiple sensors (temperature, humidity, brightness and motion, door and window sensors) and actors (light bulbs, thermostats). To make the individual setups differ from each other and evaluate the ability of \ourname to generalize, each setup also had some additional sensors and actors, making the dataset heterogeneous. For example, in the setup \datasetHomeOne also a CO$_2$ sensor was installed, while in the setups \datasetHomeFour and \datasetHomeFive also a number of smart thermostats were installed. In App.~\ref{app:devices}, we show the deployed sensors and actors for each setup. For the data collection, the popular open-source smart-home control system \mbox{Home Assistant was used (cf.~\sect\ref{sect:approach-implementation}).}

The devices were installed in different homes, covering a one-person room in a shared apartment (\datasetHomeOne), an one-person apartment (\datasetHomeTwo), as well as shared homes with 4 inhabitants each (\datasetHomeThree, \datasetHomeFour, \datasetHomeFive).
The experiments included ten different male and female participants (teenagers, students, and adults up to approximately 49 years). Initially, controlled experiments incorporating a number of simulated attack scenarios were executed in the simpler attack settings in \datasetHomeOne and \datasetHomeTwo, since these environments included only one inhabitant and were therefore easy to control. The more complex contextual settings incorporating several persons in \datasetHomeThree and \datasetHomeFour were used for passive data collection, without active attacks, mainly to test the sensitivity of the approach for false alarms under a more challenging setting. Finally, the most complex IoT set-up was implemented in the multi-person setting \datasetHomeFive, where also controlled experiments with attacks were implemented to test the full performance of \ourname in complex real-world settings. Each setup used the home automation platform for automatically triggering actions, e.g., turn off the camera when the user comes home, turn of the heating when the window is opened, or reduce the heating temperature during the night.

\subsubsection{Ethical Considerations}
\label{sect:eval-datasets-ethics}
The dataset collection raised ethical concerns, as the recorded behavior of the users might contain sensitive data. We addressed these concerns by ensuring that all affected persons, i.e., the users as well as all guests, were aware of the data collection and gave their consent. Further, we limited the approach to non-privacy-sensitive sensors and excluded the other sensors like the geolocation or the SSID of the WiFi network that the mobile phone is connected to. In addition, all potentially sensitive data items were anonymized. Our experimental set-up has been reviewed and approved by the ethics board of our university.

\subsection{Experimental Results}
\label{sect:eval-results}

The performance of \ourname for the individual setups is shown in Tab.~\ref{tab:eval-results-overview}. The table shows the results in terms of \fpr and for the setups where attacks were performed (\datasetHomeOne, \datasetHomeTwo, \datasetHomeFive), also the Precision (\precision), Recall (\recall), and F1-Score. As Tab.~\ref{tab:eval-results-overview} shows, \ourname recognizes almost all benign events correctly ($\text{\fpr} \leq 0.3\%$) and thereby fulfills \criteria{3} ( Minimizing false alarms). Tab.~\ref{tab:eval-results-overview} also shows that \ourname detects almost all the attacks ($\text{\recall}\geq 99.64\%$). The detailed results for the individual attacks are shown in Tab.~\ref{tab:eval-results-attacks} and discussed in~\sect\ref{sect:eval-results-attacks}.

\subsubsection{Attack Detection}
\label{sect:eval-results-attacks}

\begin{table}[tb]
\caption{Categorization of evaluated attacks into Event Spoofing (ES), Event Interception (EI), Command Spoofing (CS), and Command Interception (CI)}
\label{tab:attacks_categorization}
\centering

\begin{tabular}{l|ccccc}
\textbf{Attack} & \textbf{ES} & \textbf{EI} & \textbf{CS} & \textbf{CI} \\ \hline
Door open while absent & & \devicePresent & \devicePresent & \devicePresent \\
Lights on while absent & \devicePresent & & \devicePresent &\\
Movement while absent & \devicePresent & & &  \\
Camera off while absent & \devicePresent & & \devicePresent  &\\
Light flickering &  &  & \devicePresent & \\
Heating on while open windows & \devicePresent & \devicePresent &  & \devicePresent\\
Lights on during night & & & \devicePresent &  \\
Fake Fire closed windows & \devicePresent & & &  \\
Fake Fire open windows & \devicePresent & & & 
\end{tabular}
\end{table}

To evaluate the performance of \ourname, we performed various attacks. The malicious behavior for these attacks can be categorized as follows:\\
1) \textbf{Event Interception (EI)}: \adversary intercepts (i.e., suppresses) an event, for instance,  \adversary intercepts the event "\texttt{window open}" to not turn off the heating, creating a potential monetary damage to the user.\\
2) \textbf{Event Spoofing (ES)}: \adversary spoofs (i.e., creates) a fake event. For instance, \adversary creates the fake event "\texttt{user at home}" when the user is not at home to trigger the command "\texttt{turn off the camera}", e.g., for breaking into the house without being recorded.\\
3) \textbf{Command Interception (CI)}: \adversary intercepts (i.e., suppresses) a command. For instance, when the user is leaving and triggering the "\texttt{lock the door}" command, \adversary can maliciously intercept the command leaving the door open.\\
4) \textbf{Command Spoofing (CS)}: \adversary spoofs (i.e., creates) a fake command. For instance, \adversary can trigger fake commands, e.g., to turn on the light while the user is sleeping or absent, creating a potential damage (device damage and/or electricity costs) to the user.

As Tab.~\ref{tab:attacks_categorization} shows, some attacks such as "Door open while absent" may actually fall into  multiple categories: i) EI: when the user is leaving it is automatically propagated the event "\texttt{user not at home}". \adversary can intercept the event avoiding that the apartment door is locked. ii) CS: when the user is not at home, \adversary can spoof the command "\texttt{open the door}", having physical access to the home. iii) CI: when the user leaves, \adversary can intercept the command "\texttt{lock the door}", preventing it to be locked.

\begin{table}[t]
	\centering
	\caption{Performance of \ourname on Real-World Attacks, all values in percentage.}
	\label{tab:eval-results-attacks}
        \scalebox{0.85}{  
		\begin{tabular}{l|rrrrrr}
			\multicolumn{1}{c|}{Attack} &\multicolumn{1}{c}{Dataset} & \multicolumn{1}{c}{\precision} & \multicolumn{1}{c}{\recall} & \multicolumn{1}{c}{\fScore}\\\hline
			
			\multirow{2}{4cm}{Door Open During Absence} & \datasetHomeOne & 100.0 & 100.0 & 100.0\\
			& \datasetHomeFour&  100.0 & 100.0 & 100.0\\
			\hline
			
			\multirow{3}{4cm}{Lights On During Absence} & \datasetHomeOne & 100.0 & 100.0 & 100.0\\
			& \datasetHomeTwo & 100.0 & 100.0 & 100.0 \\
			& \datasetHomeFour&  100.0 & 100.0 & 100.0\\
			\hline
			
			\multirow{2}{4cm}{Movement During Absence} & \datasetHomeTwo & 100.0 & 100.0 & 100.0\\
			& \datasetHomeFour & 100.0 & 100.0 & 100.0\\\hline
			
			\multirow{5}{4cm}{Light Flickering}  & \datasetHomeOne & 100.0 & 100.0 & 100.0\\
			& \datasetHomeTwo & 100.0 & 100.0 & 100.0 \\
			& \datasetHomeThree & 100.0 & 100.0 & 100.0 \\
			& \datasetHomeFour & 100.0 & 100.0 & 100.0 \\
			& \datasetHomeFive & 100.0 & 100.0 & 100.0 \\\hline
			
			\multirow{3}{4cm}{Heating while Windows Open} & \datasetHomeOne & 100.0 & 100.0 & 100.0 \\
			& \datasetHomeTwo & 100.0 & 100.0 & 100.0 \\
			& \datasetHomeFive & 100.0 & 100.0 & 100.0\\\hline
			
			\multirow{5}{4cm}{Lights On During Night} & \datasetHomeOne & 100.0 & 100.0 & 100.0\\
			& \datasetHomeTwo & 100.0 & 100.0 & 100.0 \\
			& \datasetHomeThree & 100.0 & 100.0 & 100.0 \\
			& \datasetHomeFour & 100.0 & 100.0 & 100.0 \\
			& \datasetHomeFive & 100.0 & 100.0 & 100.0\\\hline
			
			\multirow{3}{4cm}{Fake Fire Open Windows} & \datasetHomeOne & 100.0 & 100.0 & 100.0\\
			& \datasetHomeTwo & 100.0 & 100.0 & 100.0 \\
			& \datasetHomeFour&  100.0 & 100.0 & 100.0\\

			\multirow{3}{4cm}{Fake Fire Closed Windows} & \datasetHomeOne & 100.0 & 100.0 & 100.0\\
			& \datasetHomeTwo & 100.0 & 100.0 & 100.0 \\
			& \datasetHomeFour&  100.0 & 100.0 & 100.0\\

	\end{tabular}}
\end{table}

\subsubsection{Internal Components}
\label{sect:eval-results-components}
\begin{figure}[bt]
	\centering{
    	\subfloat[\datasetHomeFive]{
    	    \hspace{-0.05cm}
    		\includegraphics[clip,width=0.5\columnwidth]{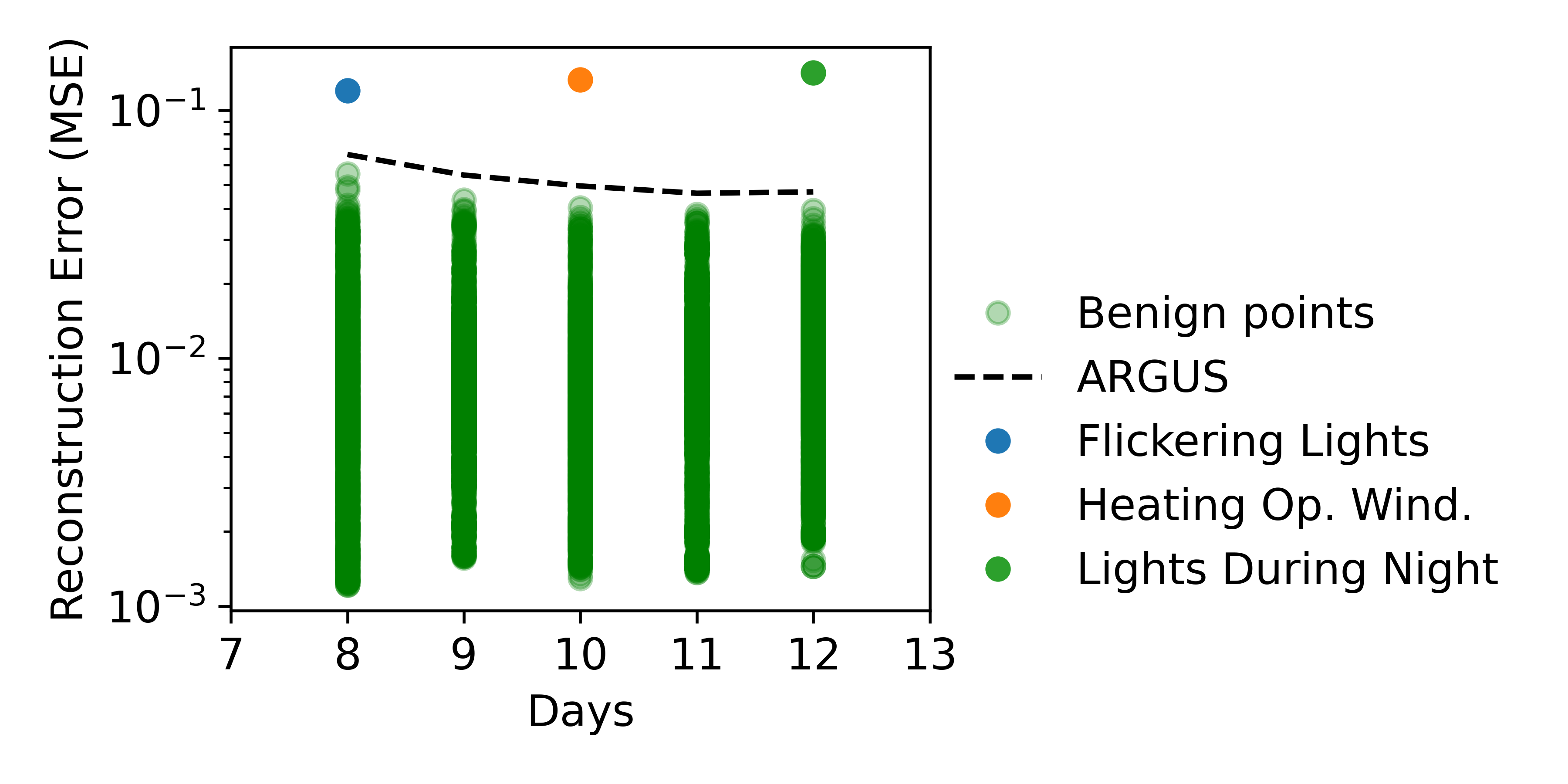}
    		\label{fig:eval-results-components:scoreScatter:home5}
    	}
    	\subfloat[\datasetHomeFour]{
    	    \hspace{-0.1cm}
    		\includegraphics[clip,width=0.5\columnwidth]{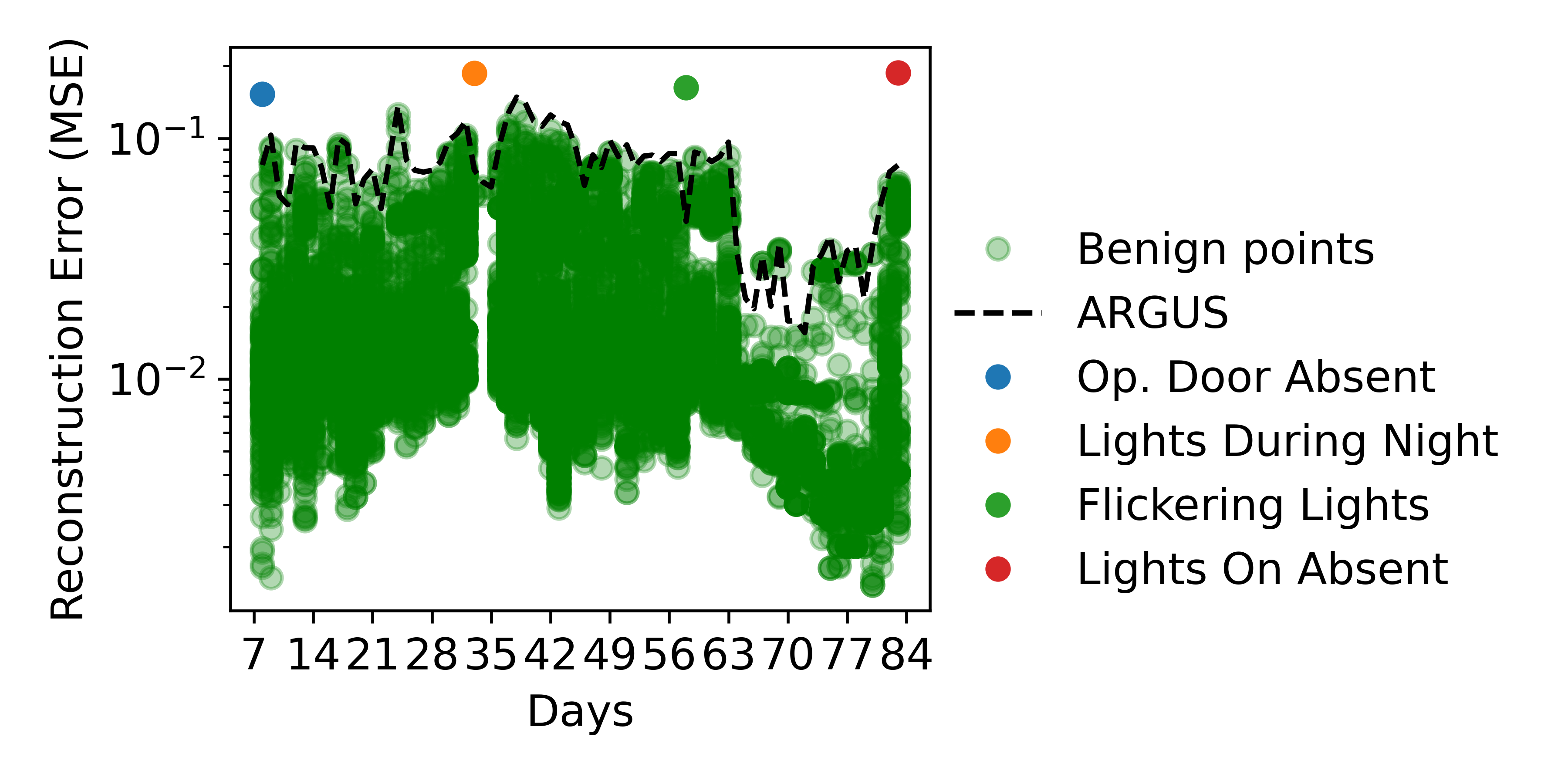}
    		\label{fig:eval-results-components:scoreScatter:home4}
    	}
    	\\
    	\subfloat[\datasetHomeOne]{\hspace{-0.1cm}
    	    \includegraphics[clip,trim={.3cm 0 0.375cm 0},width=0.985\columnwidth]{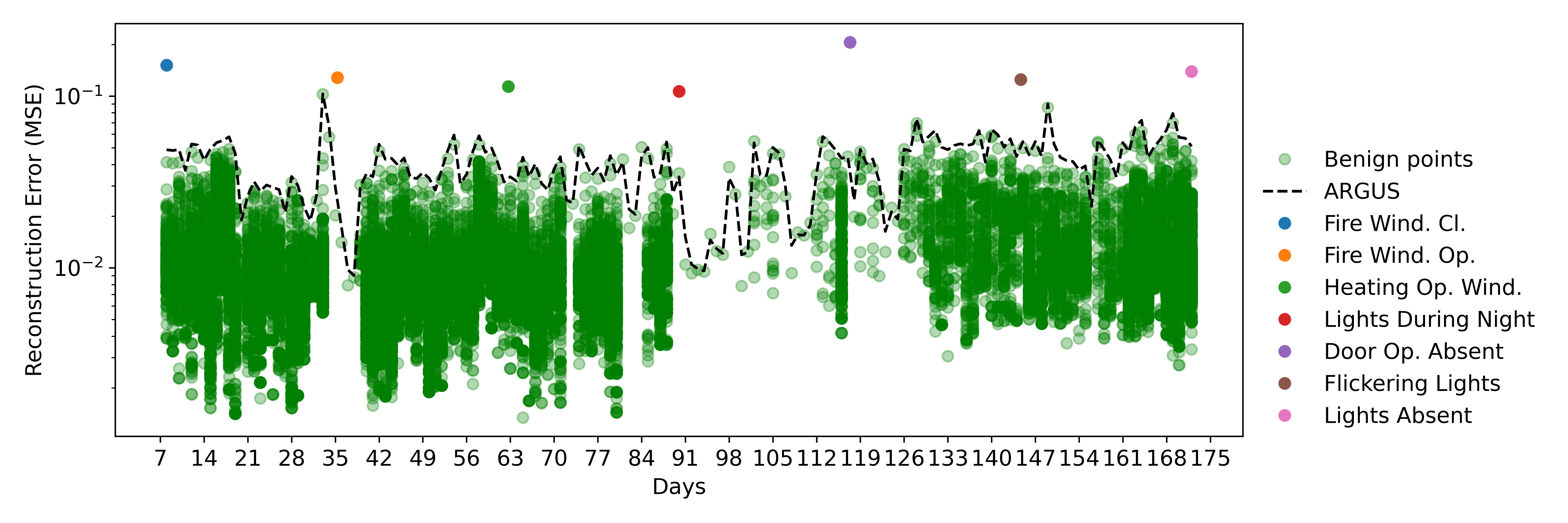}
    		\label{fig:eval-results-components:scoreScatter:home1}
    	}
    	
	}
	\caption{Anomaly Scores for different real-world homes for benign behavior and attacks.}
	\label{fig:eval-results-components:scoreScatter}
\end{figure}

\paragraph{\neuralNetworkComponent.} The \neuralNetworkComponent component predicts the reconstruction loss that is used by the \classificationComponent component to discriminate benign events and attacks. Fig.~\ref{fig:eval-results-components:scoreScatter} shows the predicted anomaly scores for benign events and the performed attacks in the setups \datasetHomeOne, \datasetHomeTwo, and \datasetHomeFive. As the figure shows, the \neuralNetworkComponent component separates the attacks well from the benign events, allowing the threshold to effectively distinguish between both event types. In App.~\ref{app:aeAlternatives} we evaluate alternative design choices for the \neuralNetworkComponent component.

\begin{figure}[bt]
	\centering
	\includegraphics[width=0.55\columnwidth]{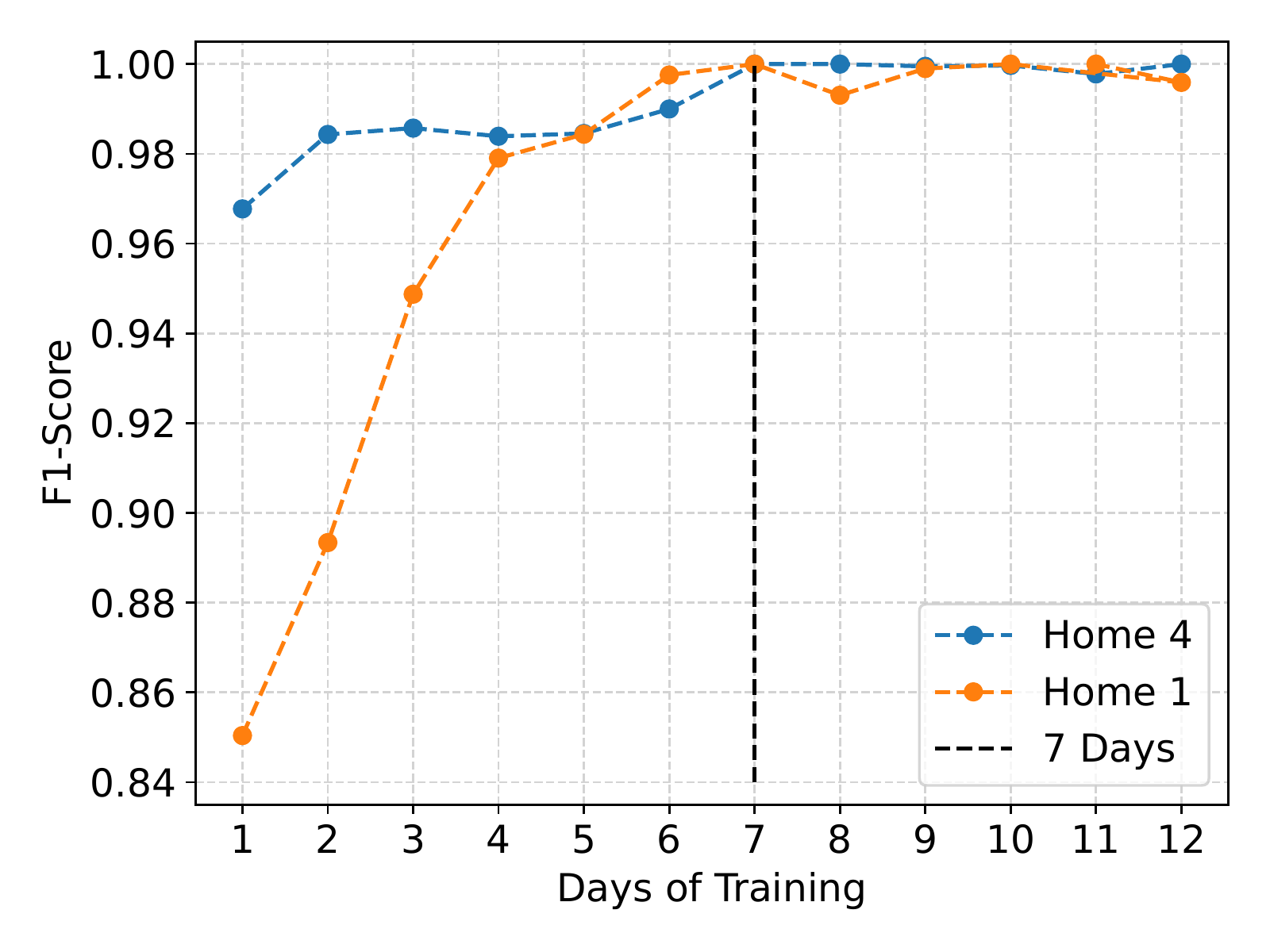}
	\caption{Evaluation of \ourname depending on the duration of training data for the setup \datasetHomeOne and \datasetHomeFour}
	\label{fig:eval-results-components:trainingData}
\end{figure}

As Subfig.~\ref{fig:eval-results-components:scoreScatter:home1} shows, 7 days of training data are sufficient for \ourname to model the expected context, such that only very few FPs are predicted ($\text{\fpr}\leq0.03\%$) even for long periods (80 days) without any adaption or retraining. This demonstrates the effectiveness of the \neuralNetworkComponent \mbox{component for modeling the user's behavior.}\\
To evaluate the amount of training data that \ourname needs to effectively model the user's behavior, we performed multiple experiments with different duration for capturing the training data in the setup \datasetHomeOne. As Fig.~\ref{fig:eval-results-components:trainingData} shows, 7 days of training data are sufficient for \ourname to achieve a F1-Score of 99.64\% on 80 days of test data. This demonstrates that \ourname is able to learn users' behavior without requiring a long training data collection phase.

\paragraph{\classificationComponent.} The \classificationComponent component uses the dynamically determined threshold (cf.~\sect\ref{sect:approach-threshold}) for recognizing attack events based on their anomaly scores.

Tab.~\ref{tab:eval-results-components:threshold} compares the proposed threshold tuning approach with different alternative options. The threshold "Mean of max" realizes a static threshold, calculated as the average of the per-day maximal anomaly scores for the training data. However, it causes FPs, such that the precision is only approx. $37.5\%$ and the threshold cannot adapt dynamically. Also, using the mean plus standard derivation of the previous day causes many FPs ($\text{\fpr}=5.9\%$) and as well as using the maximal anomaly score of the previous day as threshold or (Pr=$87.8\%$). We also evaluated different choices for the aging factor $\alpha$. For $\alpha=0.0$, resulting in using only the threshold candidate $C_d$, although no FPs were caused here, the threshold reacted too much on changing behavior, such that it was not able to detect all attacks ($\text{\recall}=99\%$). On the other side, a too high value for $\alpha$, resulting in a static threshold also failed to detect all attacks since the anomaly scores were higher than normal on the first day ($\text{\precision}=94.7\%$). In comparison, \ourname, setting $\alpha=0.2$, performed best, achieving $\text{\recall}=100.0\%$, $\text{\fpr}=0.0\%$ and $\text{\fScore}=100.0\%$. In App.~\ref{app:thresholdCandidateComparison} we evaluate the difference between the threshold and threshold candidate in detail, in App.~\ref{app:alphabeta} we evaluate different choices for the $\alpha$ and security level $\beta$.

\section{Security Considerations}
\label{sect:secConsideration}
The main goal of \ourname is to detect contextual attacks and prevent adversary \adversary from abusing the deployed IoT devices.  As we demonstrated in \sect\ref{sect:eval}, \ourname effectively detects abnormal behavior. However, if adversary \adversary is aware of the presence of \ourname, it may attempt to manipulate the system in a way that it can avoid detection while still being able to successfully execute the attacks. 

\subsection{Avoidance by Context Manipulation}
Since \ourname detects malicious actions based on the context of the action taking place, \adversary may attempt to spoof the contextual status updates of specific devices in a way that would make the actions falsely seem to happen in a legitimate context (e.g., \adversary spoofs the window sensor state as being 'closed' whereas in reality the window is open and subsequently turns on the heating to waste energy). However, this would make the attack significantly more challenging for the adversary, since it would not be sufficient to compromise the control plane of the targeted IoT devices but it would be necessary to compromise and run code on specific devices in order to impersonate them or spoof context updates. As discussed in \sect\ref{sect:design-trust}, we consider such attacks to be outside of the scope of \ourname, since there already are numerous state-of-the-art approaches for detection of direct attacks against IoT devices that are orthogonal to \ourname and can complement its detection capabilities with regard to such attacks.

\subsection{Manipulating Model Training Data}
\label{sect:secConsideration-datapoisoning}
If adversary \adversary is present in the targeted network already during the initial training phase of the system, it can trivially execute any attacks and remain undetected. In accordance with similar work, we therefore make the assumption that the system is not compromised during the initial training of the system (cf. \sect~\ref{sect:design-trust}).
However, we also evaluated a setting in which the adversary manages to start a data poisoning attack when the initial training has not been entirely completed, or, the attack is executed during a re-training phase of the detection model. We evaluated the impact of small amounts of attack data in the training datasets to examine the ability of \ourname to detect attacks (cf. App.~\ref{sect:eval-results-poisoning}), showing that \ourname is resilient against the evaluated attack.

\section{Discussion}
\label{sect:discussion}
We presented \ourname that detects contextual attacks. We demonstrated in \sect\ref{sect:eval} that it effectively detects attacks without raising a significant number of false positives ($\text{\fpr}\leq 0.03\%$) even in complex scenarios, such as homes with multiple inhabitants.
In \sect\ref{sect:problem-requirements} we discussed different requirements that an effective intrusion detection system needs to fulfill. We showed in \sect\ref{sect:approach} that \ourname fulfills our requirements with regard to fast detection (\criteria{1}), cause identification (\criteria{2}), and autonomous operation (\criteria{4}). Our evaluation in \sect\ref{sect:eval-results} also showed that \ourname generates a very low number of false alarms, thereby satisfying requirement \criteria{3}.
\subsection{Limitations of \ourname}
\label{sect:discussion-limits}

\begin{table}[tb]
	\centering
	\caption{Evaluation of Different Choices for the Classification Threshold for the setup \datasetHomeFive, all values in percentage.}
	\label{tab:eval-results-components:threshold}
		\begin{tabular}{l|rrrr}
			\multicolumn{1}{c|}{Threshold} &\multicolumn{1}{c}{\fpr} & \multicolumn{1}{c}{Pr} & \multicolumn{1}{c}{Re} & \multicolumn{1}{c}{F1-Score} \\\hline			
			Mean of max & 0.1 & 37.5 & 100.0 & 54.5\\
			Mean+Std of prev. day & 5.9 & 0.7 & 100.0 & 1.4\\
			Max of prev. days & 0.1 & 87.8 & 100.0 & 93.5\\\hline
			\ourname ($\alpha=0.0)$&  0.0 & 99.0 & 99.0 & 99.9\\
			\ourname ($\alpha=1.0)$&  0.002 & 94.7 & 99.9 & 97.3 \\\hline
			\ourname & 0.0 & 100.0 & 100.0 & 100.0
	\end{tabular}
\end{table}
The focus of \ourname is to recognize behavior that deviates from the normal behavior of the legitimate user. In \sect\ref{sect:eval} we demonstrated that \ourname can also handle very complex scenarios like a home with multiple inhabitants. However, if the change in the normal behavior is too large to be handled by \ourname, e.g., when additional persons move into the apartment, a retraining of the model must be initiated. This could be done automatically if the user labels raised alerts as false positives (e.g., using their smartphone). If the False-Positive-Rate is too high, a retraining can be initiated automatically. Another option would be to continuously adapt the model slightly, according to the monitored behavior. We leave determining suitable strategies for this to future work.

Another limitation is that, since \ournameGen definition of anomalous behavior depends on the behavior of the user during training time, actions that the user performed frequently during the training time might not be detected. This might be a problem if the user wants to use \ourname, to unlearn bad habits, e.g., learning to turn of the heating when the window is opened. As this action is performed by the user, using \ourname to change the user's behavior is out-of-scope of this paper. To handle those scenarios, \ourname could be extended by a policy-based detection component. As discussed in \sect\ref{sect:intro}, this policy-based detection is not suitable to detect attacks in general but would be able to detect a single, specific, previously defined situation (like an active thermostat while the windows are open).

Further, \ourname focuses on observing the state of the system to recognize illegitimate states. Therefore, it can be seen as complementary to policy-based approaches, that, e.g., supervise the execution of critical automation rules in few, well-defined situations. An example here would be an automation rule that locks the door exactly 10 minutes after the user left the house. Since this behavior does not depend on other circumstances but only on the time since the user left the home, it can be easily supervised by a manually crafted policy or \sota policy-based approaches. In comparison, \ourname focuses on complex situations. For example, turning on the light depends on many different contextual factors, e.g., the presence of the user, whether he is sleeping, etc.

Moreover, since \ourname uses the observed behavior for modeling the expected behavior, \ourname is limited to detecting abnormal situations but cannot supervise the behavior of IoT devices to handle these abnormal situations. An example would be an automation rule that automatically notifies the user, when a flood sensor notifies water (cf. work of Celik \etal~\cite{celik2019iotguard}). Since the flooding situation is unlikely to occur in the training data, \ourname cannot detect if the attacker suppresses the sending-notification action, as suppressing actions is outside of our adversary model (cf.~\sect\ref{sect:design-trust}). However, since as the flooding situation is already anomalous itself, \ourname will notify the user about the flooding, such that the user can take effective counter measures himself.

\begin{table*}[t]
	\centering
	\caption{Comparison of Approaches for Contextual Anomaly Detection. The symbol indicates the presence (\presentSign) or absence (\absenceSign) of the respective ability, while the color indicates whether the presence/absence is desired (green) or undesired (red).}
\scalebox{0.8}{  
	\label{tab:related-work}
		\begin{tabular}{l|ccccccccc}
			Approach & \specialcell{Restricted\\ to known\\ Attacks} & \specialcell{Bound to\\ Policies} & \specialcell{Considers\\ Normal\\ User Behavior} & \specialcell{Handles\\ unknown\\ Devices} & \specialcell{Infers\\Hidden\\Correllations} &\specialcell{Handles \\Event\\ Spoofing} & \specialcell{Handles\\ Event\\ Interception} & \specialcell{Handles\\ Command\\ Spoofing} & \specialcell{Handles\\ Command\\Interception} \\
			\hline\\
			Amraoui \etal~\cite{amraoui2020ml} & \absentGood & \absentGood & \presentGood & \presentGood & \absentBad & \absentBad & \absentBad & \presentGood & \absentBad\\
			HomeGuardian~\cite{dai2022homeguardian} & \presentBad & \absentGood & \presentGood & \presentGood & \presentGood & \presentGood & \presentGood & \presentGood & \presentGood \\
			HAWatcher~\cite{fu2021hawatcher} & \absentGood & \presentBad & \presentGood & \absentBad & \presentGood&  \presentGood & \presentGood & \presentGood & \presentGood\\
			6thSense~\cite{sikder20176thsense} & \absentGood & \absentGood & \presentGood & \presentGood & \absentBad & \presentGood & \presentGood & \presentGood & \presentGood \\
			Aegis~\cite{sikder2019aegis} & \absentGood & \absentGood & \presentGood & \presentGood & \absentBad & \presentGood & \presentGood & \presentGood & \presentGood \\
			Tang \etal~\cite{tang2019smart} &\presentBad & \absentGood & \presentGood & \presentGood & \presentGood & \presentGood & \presentGood & \presentGood & \presentGood\\
			\hline
			
			\textbf{\ourname} & \absentGood & \absentGood & \presentGood & \presentGood & \presentGood & \presentGood & \presentGood & \presentGood & \presentGood 
	\end{tabular}}
\end{table*}

\subsection{False-Alerts}
\label{sect:discussion-fp}
\ourname only creates very few False-Positives (FPs) as it is 0.03\% for \datasetHomeOne, and 0\% for all of the other homes (\datasetHomeTwo, \datasetHomeThree, \datasetHomeFour, \datasetHomeFive). Thus, even for a high number of daily events (e.g., 3000), this would result only in a single FP per day. Considering the high number of notifications that users receive on their smartphones each day, e.g., socials and webservices as Google, one false alert per day is negligible.

\section{Related Work}
\label{sect:sota}
There is a large body of literature on IoT security with many diverse approaches, although not all are relevant for our paper. We categorized them as follows: i) Network Traffic Inspection, ii)  Command Authentication, iii) Data Provenance, iv) Local Intrusion Detection, v) Sensor Validation, vi) Policies \& Transition Graphs, vii) Contextual Anomaly Detection. 

Approaches that inspect the network traffic~\cite{homesnitch,diot,fan2020iotdefender} are not effective for detecting contextual attacks that compromise the control plane. Since here, the attacker activate the functions of the IoT devices over the regular control infrastructure, the network packets that are used for transmitting commands are indistinguishable from the traffic of benign commands. By analyzing the context of the invocation, i.e., states of other devices, \ourname can detect contextual attacks.

Command Authentication approaches~\cite{tian2017smartauth,celik2019iotguard,JiaNDSS2017ContexIOT} focus on apps that are responsible for automating various tasks. Malicious apps could try to activate functions on devices which they are not supposed to control, e.g., an app that is responsible for measuring the humidity could try to unlock the door. To prevent such attacks, these approaches authenticate the source of a triggered command and check if the respective apps is actually allowed to control the targeted device. However, compared to \ourname they cannot detect attacks, where the attacker uses the regular control infrastructure, e.g., use the cloud service of the smart-lock to unlock the door.

Local intrusion detection approaches need to be installed on each IoT device locally~\cite{arshad2020intrusion,golomb2018ciota,raza2013svelte}. However, these approaches assume that the device manufacturer allows and supports installing software on the devices and that low-performance devices can execute additional software, while, \ourname does not require any modification of the devices.

Data Provenance Approaches analyze the source of a command. One example for approaches falling into this category is ProvThings, being developed to explain attacks in the retrospective. By analyzing the flow of actions that were automatically triggered, the attack helps especially to explain complex attacks, where the attacker did not trigger the targeted action directly but exploited and concatenated multiple automation rules~\cite{rahman2018fear}. However, since this scheme focuses on explaining a performed attack rather than detecting the attack while it is being performed, ProvThings is complementary to \ourname.

Hence, in the rest of this section, we focus on those proposals (iv-vii) that consider the whole \emph{context} of the underlying IoT system in various ways such as validating the individual sensor values~\cite{adkisson2021autoencoder,yasaei2020iot,yin2020anomaly,kotevska2019kensor}, using policies or modeling the system states as nodes of a graph and analyzing the state transitions~\cite{yamauchi2020anomaly,feng2019systematic,yahyazadeh2019expat,celik2018soteria,choi2018detecting,kafle2021towards}, or focusing on detecting contextual anomalies~\cite{fu2021hawatcher,amraoui2020ml,tang2019smart,sikder2019aegis,sikder20176thsense}.

\subsection{Sensor Value Validation}
\label{sect:sota-sensors}
Adkisson~\etal use auto-encoders to detect anomalous values of sensors that measure the environment for plants~\cite{adkisson2021autoencoder}.
Kotevska~\etal compare the values of co-located sensors~\cite{kotevska2019kensor}. Yasaei~\etal compare the values of different sensors where the measured values directly effect each other, e.g., the water temperature and the Nitrate concentration, to detect malfunctions of individual sensors~\cite{yasaei2020iot}. 
Yin~\etal analyze time series of the individual sensor values~\cite{yin2020anomaly}. 

However, all these approaches consider only simple scenarios without human-caused randomness or are limited to detecting sensor malfunctions without being able to detect anomalous system states that occur in the real world. For example, if the door of an apartment is opened while the inhabitant is absent, then these approaches will not create an alert, as the co-located or correlated sensors consistently show that the door is opened. In comparison, \ourname can detect that the door is opened in the wrong context (while the inhabitant is not at home) and inform the user.

\subsection{Policies \& Transition Graphs}
Other approaches verify the system state against pre-defined policies~\cite{kafle2021towards,yahyazadeh2019expat,yamauchi2020anomaly}. However, these rules can easily be analyzed by the adversary and also create additional overhead for the user to define these policies. For example, for HomeEndorser~\cite{kafle2021towards} and Expat~\cite{yahyazadeh2019expat} the policies have to be defined manually, making the system unpractical. Yamauchi \etal monitor the users' behavior during the setup time and create an alert if the performed action was not invoked during the setup phase, in a situation where all sensors measured the exact values as they do at the moment~\cite{yamauchi2020anomaly}. Feng \etal automatically craft invariants from logfiles of industrial IoT systems~\cite{feng2019systematic}, which are significantly less complex then smart-homes and can only learn linear relations between different devices, increasing the risk of wrong predictions. In comparison to these approaches, \ourname is trained automatically from the data that is captured during the setup phase. The deep learning-based behavior modeling of \ourname enables it to learn hidden/non-linear relationships between system states, e.g., it does not require all sensors to measure the same states as in the setup phase but learns which sensors can actually differ (e.g., temperature sensors) and which sensor values must be exactly the same (e.g., the presence of the inhabitant).

Homonit~\cite{zhang2018homonit} and Soteria~\cite{celik2018soteria} both extract a Deterministic Finite Automaton (DFA) from the source code of mobile apps, where each node represents a system state. They create an alert if an invalid transition from one state to another state is taken. However, since these approaches use only the source code as data source, they consider only a theoretical legitimate behavior and not the actual user's behavior for recognizing contextual attacks. For example, the app might allow to turn on the light while the user is sleeping. However, if the user does not do this, then \ourname can still detect this attack. DICE uses a Markov Chain to model the transitions of a system~\cite{choi2018detecting}. However, it models each system state as all sensor values that occurred in a time frame of 1 minute, such that it cannot distinguish, e.g., between turning on the light or the light-flickering attack that was discussed in \sect\ref{sect:eval-results-attacks}.

\subsection{Contextual Anomaly Detection}
Table~\ref{tab:related-work} shows on overview of \sota approaches for Contextual Anomaly Detection. It systematize  them based on different criteria, 1) if the system can only recognize attacks that were known at setup time, 2), whether the system uses policies for recognition and is restricted to detect behavior that is allowed/forbidden by these policies, 3) whether the actual user behavior is considered or only expected behavior that is, e.g., extracted from app descriptions, 4) whether custom IoT devices, where no semantic information are available can be handled, 5) if also hidden relationships can be learned or if the system is restricted to recognizing state changes that have been exactly occurred in the training data, 6) if the system can handle attacks that spoof events (e.g., fake temperature sensor values), 7) if event interceptions are recognized (e.g., intercepting the user-coming-home event before the door is opened), 8) if spoofed commands can be recognized (e.g., if the light is turned on during the night), and 9) if command interceptions are recognized (e.g., the command to turn of the heating before opening the window).

HAWatcher uses semantic information, extracted from the app descriptions and source code, to match devices with the correlated attribute. For example, it determines that a sound sensor and a media player both are connected via the sound property. HAWatcher then generates correlation rules that need to be fulfilled~\cite{fu2021hawatcher}. This structure enables HAWatcher to enforce strict policies (cf.~\sect\ref{sect:discussion-limits}). However, requiring the presence of semantic information limits the applicability as such information are not always available. For instance, the temperature, humidity, and CO$_2$ sensor in our experimental setup were built by the respective inhabitants. Therefore, for these devices neither apps nor semantic information exists. In comparison, \ourname uses only the actually occurred sensor values for learning the expected behavior, such that it is also effective without any semantic information.

Amraoui \etal train a One-class SVM for analyzing the triggered commands but do not consider the values of the sensors~\cite{amraoui2020ml}. Thus, it cannot detect, e.g., the Movement During Absence attack.

Tang \etal provide an approach that trains binary classifiers on legitimate data as well as attack data to detect sensor failures~\cite{tang2019smart}. Dai \etal use a Neural Network that is trained on attacks and benign behavior to distinguish between them~\cite{dai2022homeguardian}. However, by requiring attack training data these approaches are restricted to detecting known attacks, while \ourname can detect arbitrary attacks.

6thSense~\cite{sikder20176thsense} and Aegis~\cite{sikder2019aegis} both consider the changes of the system states (e.g., sensor updates) as Markov Chains and require the probability for invalid state transitions, i.e., attacks, to be 0. However, the system then can only consider transitions that have occurred during the training phase and fails to learn hidden correlations/uncorrelations between sensor values, i.e., that some devices always change together or are unrelated, e.g., if they are located in different rooms. In comparison, the Behavior Modeling of \ourname can learn relations, e.g., the independence of two sensors.

\section{Conclusion}
The number of devices being part of the Internet of Things is increasing rapidly, while at the same time these devices control more and more functions being part in our daily life. These factors make IoT devices an attractive goal for attacks that abuse these devices to cause financial damage or harm the users. We proposed \ourname, the first dynamic system that can detect contextual attacks on connected device settings. It consists of a data collection component, monitoring devices even in a very heterogeneous landscape with devices from various manufacturers using different communication technologies and protocols. The designed Deep Neural Network in combination with the proposed dynamic threshold tuning scheme allows to distinguish between the expected user behavior and contextual attacks.

We extensively evaluated \ourname for \numSetups different settings and showed that it effectively detects contextual attacks on IoT devices while raising only few false positives ($\text{\fpr}\leq 0.03\%$), even in complex settings.

\section*{Acknowledgments}
This work was funded in part by Intel as part of the Private AI center, HMWK within ATHENE project, and the European Union's Horizon 2020 research and innovation program under grant agreement No. 952697 (ASSURED).
\def\UrlBreaks{\do\/\do-}
\bibliographystyle{plain}
\bibliography{main}

\appendix

\section{Devices}
\label{app:devices}
Table~\ref{tab:device_datasets_table} shows for each setup in the collected dataset, a detailed list of the deployed IoT devices and measured values. In our experiments, we leveraged all values that were provided by the used home automation platform, covering the different categories of contextual features (cf.~\sect\ref{sect:problem-system}): i) Sensors/devices that measure ambient or temporal features (e.g., temperature, humidity, and luminosity), ii) user features (e.g., user presence and sleep confidence), and event features (e.g., states of the light bulbs, doors or windows).
\section{Runtime Performance}
\label{app:runtime}
The real-time approach evaluates events without delay as soon as they are captured. Compared to the time the user needs to react manually, even an unlikely delay of multiple seconds would be negligible.\\
We performed additional experiments to evaluate \ourname on a low-performance device (Raspberry Pi) without observing any delay. Further, we also restrict the main memory that \ourname was allowed to use and observed that 330MB are sufficient for the implementation. It should be noted that the prototype of \ourname was not optimized for runtime performance, such that by using, e.g., more efficient languages the runtime performance could be increased even further.

\begin{table}[th!]

\caption{Deployed devices in the collected real-world IoT dataset. The deployment of a sensor/actor is indicated by \devicePresent, while the absence is indicated by \deviceNotPresent.}
\label{tab:device_datasets_table}
\scalebox{0.65}{  
\begin{tabular}{l|ccccc}
Device & \rotatebox[origin=c]{90}{\mbox{ }\datasetHomeOne} & \rotatebox[origin=c]{90}{\mbox{ }\datasetHomeTwo} & \rotatebox[origin=c]{90}{\mbox{ }\datasetHomeThree} & \rotatebox[origin=c]{90}{\mbox{ }\datasetHomeFour} & \rotatebox[origin=c]{90}{\mbox{ }\datasetHomeFive}\\ \hline

Automation - All lights off & \deviceNotPresent & \deviceNotPresent & \deviceNotPresent & \deviceNotPresent & \devicePresent \\
Automation - All lights on & \deviceNotPresent & \deviceNotPresent & \deviceNotPresent & \deviceNotPresent & \devicePresent \\
Automation - Camera off when at home & \deviceNotPresent & \deviceNotPresent & \deviceNotPresent & \devicePresent & \deviceNotPresent \\
Automation - Dinner lights & \deviceNotPresent & \deviceNotPresent & \deviceNotPresent & \devicePresent & \devicePresent \\
Automation - Dinner table light & \deviceNotPresent & \deviceNotPresent & \deviceNotPresent & \deviceNotPresent & \devicePresent \\
Automation - Gaming mode & \deviceNotPresent & \deviceNotPresent & \deviceNotPresent & \deviceNotPresent & \devicePresent \\
Automation - Heating boost off & \deviceNotPresent & \deviceNotPresent & \deviceNotPresent & \deviceNotPresent & \devicePresent \\
Automation - Light off when no motion & \deviceNotPresent & \deviceNotPresent & \deviceNotPresent & \devicePresent & \devicePresent \\
Automation - Lights off in the evening & \deviceNotPresent & \devicePresent & \deviceNotPresent & \deviceNotPresent & \deviceNotPresent \\
Automation - Lights off when too bright & \deviceNotPresent & \devicePresent & \deviceNotPresent & \devicePresent & \deviceNotPresent \\
Automation - Lights on in the morning & \deviceNotPresent & \devicePresent & \deviceNotPresent & \deviceNotPresent & \deviceNotPresent \\
Automation - Lights on when motion detected & \deviceNotPresent & \devicePresent & \deviceNotPresent & \devicePresent & \deviceNotPresent \\
Automation - Piano Light & \deviceNotPresent & \deviceNotPresent & \deviceNotPresent & \deviceNotPresent & \devicePresent \\
Automation - Sofa Lamp & \deviceNotPresent & \deviceNotPresent & \deviceNotPresent & \deviceNotPresent & \devicePresent \\
Automation - Studio Light off & \deviceNotPresent & \deviceNotPresent & \deviceNotPresent & \deviceNotPresent & \devicePresent \\
Automation - Studio Light on when motion & \deviceNotPresent & \deviceNotPresent & \deviceNotPresent & \deviceNotPresent & \devicePresent \\
Automation: Camera on when user leave & \deviceNotPresent & \deviceNotPresent & \deviceNotPresent & \devicePresent & \deviceNotPresent \\
Camera Status Sensor & \deviceNotPresent & \devicePresent & \devicePresent & \deviceNotPresent & \deviceNotPresent \\
Climate - Control access point 1 & \deviceNotPresent & \deviceNotPresent & \deviceNotPresent & \deviceNotPresent & \devicePresent \\
$\text{CO}_2$ Sensor Status & \devicePresent & \deviceNotPresent & \deviceNotPresent & \deviceNotPresent & \deviceNotPresent \\
$\text{CO}_2$ Sensor & \devicePresent & \deviceNotPresent & \deviceNotPresent & \deviceNotPresent & \deviceNotPresent \\
Control Access Room 1 Sensor & \deviceNotPresent & \deviceNotPresent & \deviceNotPresent & \devicePresent & \deviceNotPresent \\
Door Sensor & \devicePresent & \devicePresent & \devicePresent & \devicePresent & \devicePresent \\
Floor lamp & \deviceNotPresent & \deviceNotPresent & \deviceNotPresent & \deviceNotPresent & \devicePresent \\
Heating - heater valve & \deviceNotPresent & \deviceNotPresent & \deviceNotPresent & \deviceNotPresent & \devicePresent \\
Heating Temperature Sensor & \devicePresent & \devicePresent & \deviceNotPresent & \devicePresent & \deviceNotPresent \\
Homematic - Radiator Thermostat Temperature Sensor & \deviceNotPresent & \deviceNotPresent & \devicePresent & \deviceNotPresent & \deviceNotPresent \\
Humidity Sensor & \devicePresent & \devicePresent & \deviceNotPresent & \devicePresent & \devicePresent \\
IKEA Tradfri Roller Blind Sensor & \devicePresent & \deviceNotPresent & \deviceNotPresent & \deviceNotPresent & \deviceNotPresent \\
IP Camera - Light Level & \deviceNotPresent & \deviceNotPresent & \deviceNotPresent & \deviceNotPresent & \devicePresent \\
IP Camera - Motion & \deviceNotPresent & \deviceNotPresent & \deviceNotPresent & \deviceNotPresent & \devicePresent \\
IP Camera - Motion Active & \deviceNotPresent & \deviceNotPresent & \deviceNotPresent & \deviceNotPresent & \devicePresent \\
IP Camera - Pressure & \deviceNotPresent & \deviceNotPresent & \deviceNotPresent & \deviceNotPresent & \devicePresent \\
IP Camera - Sound & \deviceNotPresent & \deviceNotPresent & \deviceNotPresent & \deviceNotPresent & \devicePresent \\
Lamp consumption & \deviceNotPresent & \deviceNotPresent & \deviceNotPresent & \deviceNotPresent & \devicePresent \\
Lamp consumption (daily) & \deviceNotPresent & \deviceNotPresent & \deviceNotPresent & \deviceNotPresent & \devicePresent \\
Lamp consumption (total) & \deviceNotPresent & \deviceNotPresent & \deviceNotPresent & \deviceNotPresent & \devicePresent \\
Lamp current & \deviceNotPresent & \deviceNotPresent & \deviceNotPresent & \deviceNotPresent & \devicePresent \\
Lamp voltage & \deviceNotPresent & \deviceNotPresent & \deviceNotPresent & \deviceNotPresent & \devicePresent \\
Light - Ceiling & \devicePresent & \devicePresent & \devicePresent & \devicePresent & \devicePresent \\
Light - Desk Lamp & \devicePresent & \devicePresent & \devicePresent & \devicePresent & \devicePresent \\
Light - Living Room & \deviceNotPresent & \deviceNotPresent & \devicePresent & \deviceNotPresent & \deviceNotPresent \\
Philips Hue - Light Level Sensor 1 & \deviceNotPresent & \devicePresent & \devicePresent & \devicePresent & \devicePresent \\
Philips Hue - Light Level Sensor 2 & \deviceNotPresent & \deviceNotPresent & \devicePresent & \deviceNotPresent & \deviceNotPresent \\
Philips Hue - Motion Sensor 2 & \deviceNotPresent & \deviceNotPresent & \devicePresent & \deviceNotPresent & \deviceNotPresent \\
Philips Hue - Temperature Sensor 1 & \deviceNotPresent & \devicePresent & \devicePresent & \devicePresent & \devicePresent \\
Philips Hue - Temperature Sensor 2 & \deviceNotPresent & \deviceNotPresent & \devicePresent & \deviceNotPresent & \deviceNotPresent \\
Philips Hue - White Lamp 2 & \deviceNotPresent & \deviceNotPresent & \devicePresent & \deviceNotPresent & \deviceNotPresent \\
Philips Hue - White Lamp 3 & \deviceNotPresent & \deviceNotPresent & \devicePresent & \deviceNotPresent & \deviceNotPresent \\
Philips Hue - Motion Sensor 1 & \deviceNotPresent & \deviceNotPresent & \devicePresent & \devicePresent & \devicePresent \\
Piano lamp & \deviceNotPresent & \deviceNotPresent & \deviceNotPresent & \deviceNotPresent & \devicePresent \\
Radiator Thermostat Sensor & \deviceNotPresent & \deviceNotPresent & \deviceNotPresent & \devicePresent & \devicePresent \\
Smartphone - Battery Life & \deviceNotPresent & \devicePresent & \deviceNotPresent & \deviceNotPresent & \deviceNotPresent \\
Smartphone - Charging & \deviceNotPresent & \deviceNotPresent & \deviceNotPresent & \deviceNotPresent & \devicePresent \\
Smartphone - Charging Sensor & \devicePresent & \deviceNotPresent & \deviceNotPresent & \deviceNotPresent & \deviceNotPresent \\
Smartphone - Connected to WLAN & \deviceNotPresent & \devicePresent & \deviceNotPresent & \devicePresent & \deviceNotPresent \\
Smartphone - Detected Activity & \devicePresent & \devicePresent & \deviceNotPresent & \deviceNotPresent & \devicePresent \\
Smartphone - Light Sensor & \deviceNotPresent & \devicePresent & \deviceNotPresent & \deviceNotPresent & \deviceNotPresent \\
Smartphone - Locked & \deviceNotPresent & \deviceNotPresent & \deviceNotPresent & \deviceNotPresent & \devicePresent \\
Smartphone - Phone Status & \deviceNotPresent & \devicePresent & \deviceNotPresent & \deviceNotPresent & \deviceNotPresent \\
Smartphone - Sleep Confidence & \devicePresent & \devicePresent & \deviceNotPresent & \devicePresent & \deviceNotPresent \\
Smartphone - Sleep Segment & \deviceNotPresent & \deviceNotPresent & \deviceNotPresent & \devicePresent & \deviceNotPresent \\
Smartphone - Tracker & \devicePresent & \devicePresent & \deviceNotPresent & \devicePresent & \deviceNotPresent \\
Studio lamp & \deviceNotPresent & \deviceNotPresent & \deviceNotPresent & \deviceNotPresent & \devicePresent \\
Sun Sensor & \devicePresent & \devicePresent & \devicePresent & \devicePresent & \devicePresent \\
Temperature Sensor (ESP) & \devicePresent & \devicePresent & \deviceNotPresent & \devicePresent & \devicePresent \\
User Presence & \devicePresent & \devicePresent & \deviceNotPresent & \devicePresent & \deviceNotPresent \\
Weather - Home Location & \devicePresent & \devicePresent & \devicePresent & \devicePresent & \devicePresent \\
Weather - Town & \devicePresent & \deviceNotPresent & \devicePresent & \devicePresent & \deviceNotPresent \\
Window Sensor & \devicePresent & \devicePresent & \devicePresent & \devicePresent & \deviceNotPresent
\end{tabular}}
\end{table}~\\

\begin{figure}[bt]
	\centering
	\subfloat[\datasetHomeTwo]{\hspace{-0.2cm}
		\includegraphics[clip,trim={.2cm 0.4cm 0.35cm 0},width=0.5\columnwidth]{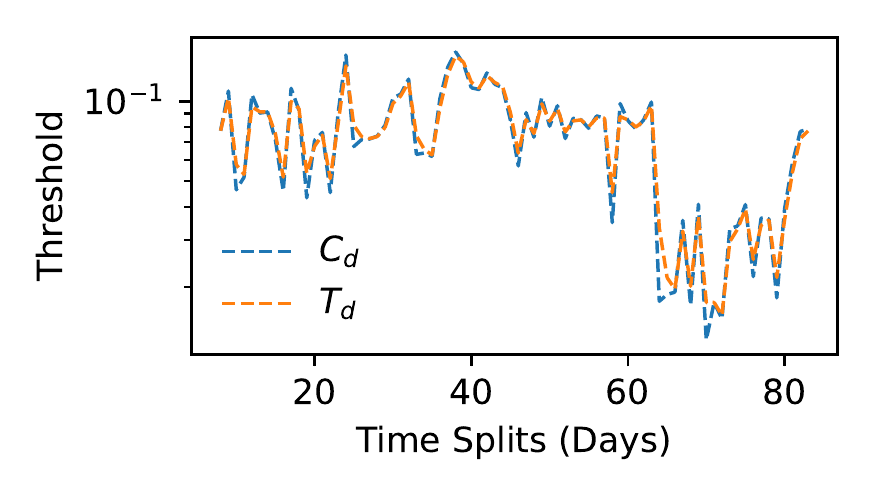}
		\label{fig:eval-results-components:candiateComparison:home4}
	}
	\subfloat[\datasetHomeFive]{\hspace{-0.2cm}
		\includegraphics[clip,trim={.2cm 0.4cm 0.35cm 0},width=0.5\columnwidth]{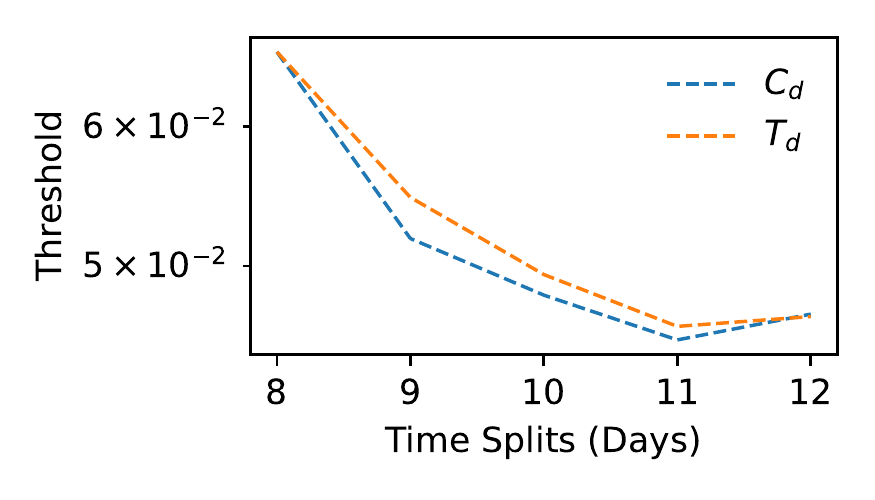}
		\label{fig:eval-results-components:candiateComparison:home5}
	}
	
	\subfloat[\datasetHomeOne]{
		\includegraphics[clip,trim={.35cm 0.4cm 0.35cm 0},width=\columnwidth]{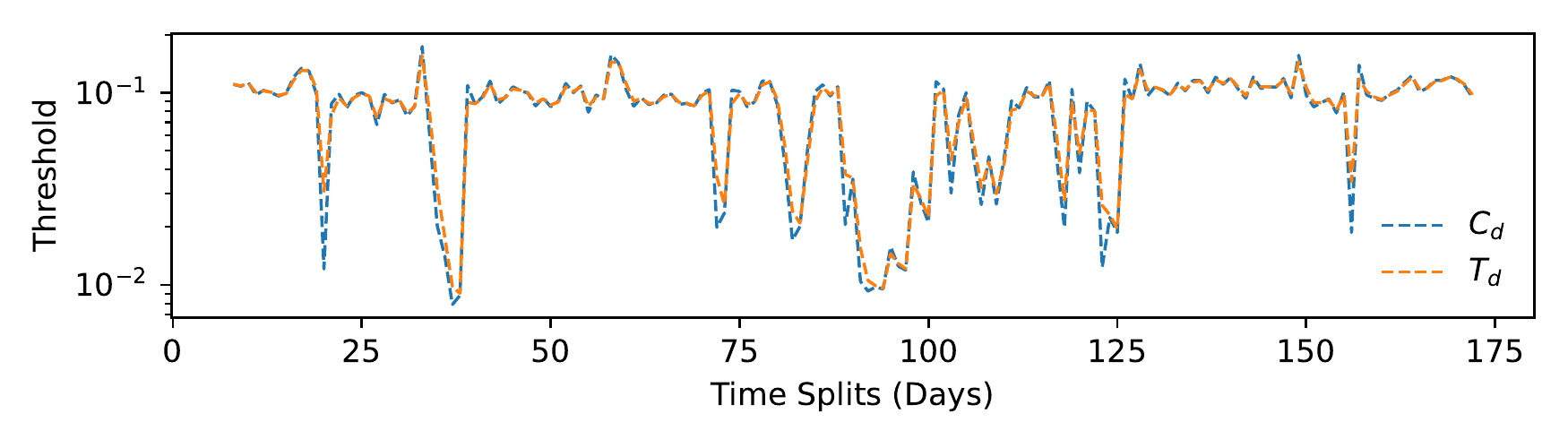}
		\label{fig:eval-results-components:candiateComparison:home1}
	}
	\caption{Thresholds values comparison between Threshold Candidate ($C_d$) and \ourname Threshold ($T_d$) for multiple real-world homes.}
	\label{fig:eval-results-components:candiateComparison}
\end{figure}

\section{Evaluation of Robustness of the Threshold}
\label{app:thresholdShifting}
In the following, we investigate whether the injection of noise in devices' states would tamper \ournameGen threshold, therefore, making it more likely to miss attacks.\\
Considering 6 different levels of noise randomly sampled from a normal distribution, described by: 
\begin{equation}
    f(x)={\frac {1}{\sigma {\sqrt {2\pi }}}}e^{-{\frac {1}{2}}\left({\frac {x-\mu }{\sigma }}\right)^{2}}
\end{equation}
where $\mu$ is the mean and $\sigma$ is the standard deviation, we attack all the devices, one per run, attacking its states with an amount of noise sampled from:\footnote{We consider the mean of 100 samples sampled from the distribution.}
\begin{equation}
f(x|\mu=1, \sigma=i) \; \forall  \; i\; \in \{1..6\}
\end{equation}
considering all the days available for testing. For understanding if the attack is successful, and with which rate, we measure the number of alerts raised and how many times the benign maximum and minimum are exceeded (i.e., if the threshold computation would be affected by the injected noise).

\begin{table}[tb]
    \centering
        \begin{tabular}{c|rrr}
            \textbf{$\sigma$} &
            \textbf{Alerts \%} &
            \textbf{Not affecting \%} &
            \textbf{Affecting no alerts \%} \\ \hline
            \textbf{1}  & 0.002  & 99.674 & 0.323    \\
            \textbf{2}  & 0.031  & 99.485 & 0.489    \\
            \textbf{3}  & 0.200  & 98.883 & 0.916    \\
            \textbf{4}  & 0.987  & 96.104 & 2.909   \\
            \textbf{5}  & 4.637  & 86.597 & 8.766   \\
            \textbf{6}  & 15.511 & 68.696 & 15.793   
        \end{tabular}%
    
\caption{Noise injection in events chain analysis on \datasetHomeOne dataset. Each entry counts 652,720$\times 18$ events evaluated. The values shown are the mean of all the devices attacked, for all the days considered per each level of noise considered (i.e., $18 \times 6$. In total are evaluated \numprint{70493760} events).}
\label{tab:threshold_shift_home1}
\end{table}

As we can observe from Table~\ref{tab:threshold_shift_home1}, in order to affect the threshold, even with just a low probability, \adversary has to inject a high amount of noise ($\sigma\in\{4,5,6\}$). However, if the level of noise gets severe the risk of raising an alert will raise up to 15.511\%. Since \adversary is assumed to be unable to compromise the \ourname system (cf.~\sect\ref{sect:design-trust}) it cannot check in advance if a noised event will have no impact, raise an alert, or affect the threshold without raising an alert. Therefore, in order to damage the detection of \ourname, it has to create a high number of attack events. Further, since the impact of a single day on the threshold is limited due to the aging factor $\alpha$, \adversary has to create many fake events over many days to damage \ourname. However, this will also cause a high number of alerts, such that the user will become suspicious and detect the attack.

\section{Comparison of Threshold and Threshold Candidate}
\label{app:thresholdCandidateComparison}
To evaluate the impact of the momentum on the threshold in more detail, we measured the threshold candidate $C_d$ as well as the resulting threshold $T_d$. As Fig.~\ref{fig:eval-results-components:candiateComparison} shows, $C_d$ fluctuates significantly, depending on the measured context of each day. However, if $C_d$ would be used as classification boundary, these fluctuations might cause FNs or FPs on the following day. For example, when the user behavior for one day $d$ differs from the usual behavior, $C_d$, which is used on the following day d+1 as classification boundary, is very high (as in Subfig.~\ref{fig:eval-results-components:candiateComparison:home1}). This would cause a risk that the anomaly scores of attacks on day d+1 are lower than $C_d$. On the other side, if the behavior of the user on day d is very similar to the expected behavior, $C_d$ might be too low, causing FPs. In comparison, the momentum used by $T_d$ smooths outliers and prevents that exceptional high or low anomaly scores on the previous day, e.g., caused by slightly differing user behavior, affect \mbox{the system's performance.}

\section{Ablation Study on $\alpha$ and $\beta$}
\label{app:alphabeta}
\begin{figure}[bt]
	\centering
	\includegraphics[width=0.75\columnwidth]{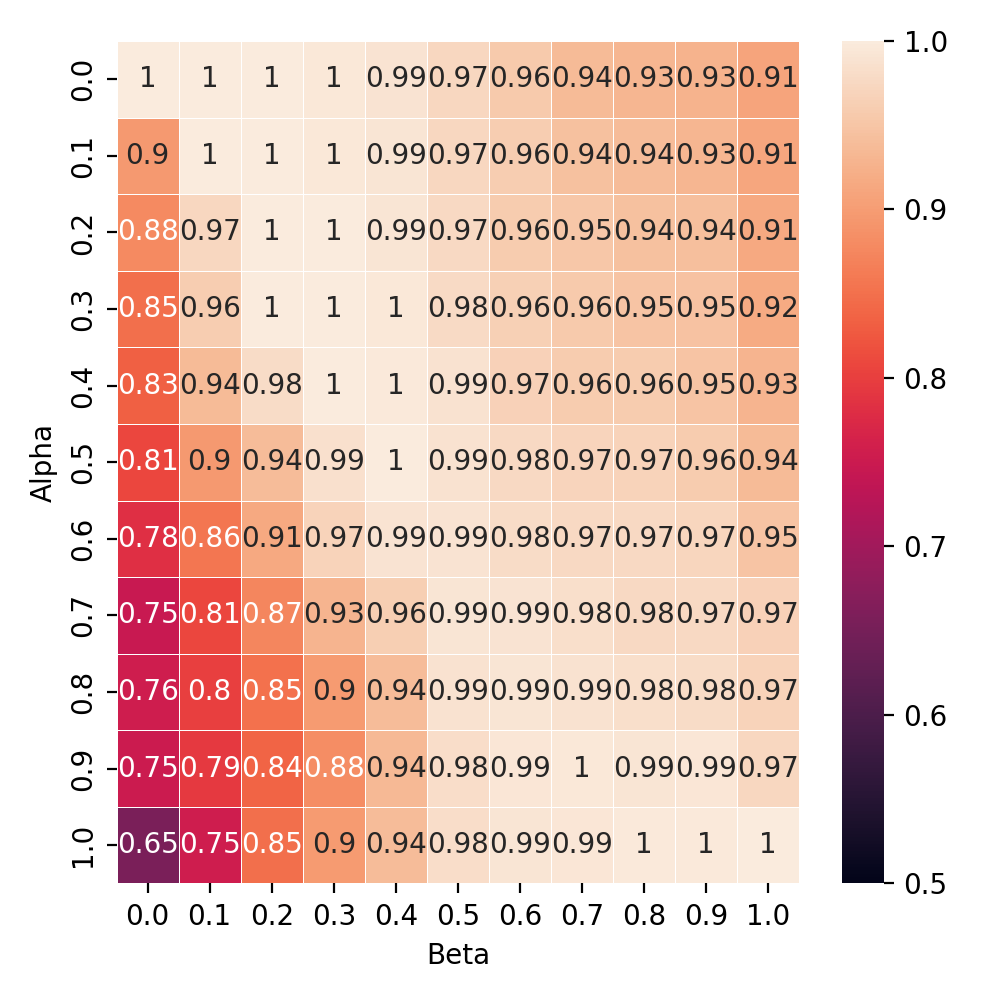}
	\caption{$\alpha$ and $\beta$ parameters selection of \ourname for the setup \datasetHomeFour. For the sake of saving space, $\alpha$ and $\beta$ are represented in the heatmap with a step of 0.1, the actual evaluation counts a step size of 0.05.}
	\label{fig:heatmap_home4_MR}
\end{figure}

Furthermore, we extensively evaluated the $\alpha$ and $\beta$ parameters selection through 400 experiments. The results are summarized in Figure~\ref{fig:heatmap_home4_MR}, showing how the selection of $\alpha=0.2$ and $\beta=0.2$ corresponds to an optimal choice.

\section{Alternatives for AE}
\label{app:aeAlternatives}
We performed multiple experiments to compare the used DNN architecture (cf.~\sect\ref{sect:approach-ml-model} to other machine learning (ML) and deep learning algorithms, the results are shown in Tab.~\ref{tab:eval-results-components:alternativeMLModels}.  As classical ML algorithm, we used a One-Class Support Vector Machine (One-Class SVM)~\cite{chen2001one}. The second approach uses also an Auto-Encoder (AE), which is also used by \ourname. However, the evaluated AE model uses normal, linear layers, while \ourname uses GRU layers to consider better the temporal order of the individual events. All approaches were trained using the same datasets and used to predict anomaly scores. However, to prevent any bias in favor of \ourname, we opted for each of the alternatives a threshold that maximizes the \fScore on the test data, while for \ourname we used the threshold that we discussed in \sect\ref{sect:approach-threshold}. However, even with this advantage, \recall is significantly lower for the One-Class SVM and the AE without the GRU layers, indicating that they miss many attacks. On the other side, especially the AE without the GRU layers also misclassifies many benign events ($\text{\fpr}=4.9\%$) and the OneClassSVM misses 25 benign events ($\text{\fpr}=0.016\%$), such that their \fScores are only 92.2\% and 68.2\%. In comparison, \ourname detects all attacks in this setup \mbox{($\text{\fpr}=0.0\%$) and the \fScore is 100.0\%}.

\begin{table}[tb]
	\centering
	\caption{Evaluation of alternative choices for the Machine Learning (ML) algorithm for modeling the expected behavior of \datasetHomeFive, all values in percentage.}
	\label{tab:eval-results-components:alternativeMLModels}
		\begin{tabular}{l|rrrr}
			\multicolumn{1}{c|}{ML Algorithm} &\multicolumn{1}{c}{\fpr} & \multicolumn{1}{c}{Pr} & \multicolumn{1}{c}{Re} & \multicolumn{1}{c}{F1-Score} \\\hline
   
		    One-Class SVM  & 0.0 & 96.2  & 52.9 & 68.2\\
			AE without GRU & 4.9 & 98.0  & 87.1 & 92.2\\\hline
			\ourname       & 0.0 & 100.0 & 100.0 & 100.0
	\end{tabular}
\end{table}

\section{Evaluation Metrics}
\label{app:metrics}
To evaluate the performance of the trained model we use common performance metrics such as False-Positive-Rate (\fpr), Precision (\precision), Recall (\recall), and \fScore. For calculating these metrics, we count the number of benign events that are correctly classified (TN) or misclassified (FN) as well as the number of attacks that are correctly recognized (TP) or not recognized (FN). 
\mbox{We define the performance metrics as follows:}\\
\textbf{False-Positive-Rate (\fpr)} indicates the risk to misclassify benign events. It is given by:
\begin{equation}
	\text{\fpr} = \frac{\text{FP}}{\text{FP} + \text{TN}}
\end{equation}
\textbf{Precision (\precision)} indicates the probability that an event that is recognized as anomaly is actually an attack. It is given by:
\begin{equation}
	\text{\precision} = \frac{\text{TP}}{\text{TP} + \text{FP}}
\end{equation}
\textbf{Recall (\recall)} indicates the effectiveness of an approach to detect attacks. It is given by:
\begin{equation}
	\text{\recall} = \frac{\text{TP}}{\text{TP} + \text{FN}}
\end{equation}
\textbf{\fScore} balances \precision and \recall. It is calculated as:
\begin{equation}
	\text{\fScore} = 2\cdot\frac{\text{\precision} \cdot \text{\recall}}{\text{\precision} + \text{\recall}}= \frac{\text{TP}}{\text{TP} + \nicefrac{1}{2}(\text{FP}\cdot \text{FN})}
\end{equation}

\section{Robustness of \ourname against Data Poisoning}
\label{sect:eval-results-poisoning}

We assume that the IoT system is not compromised during the data collection phase (cf. security assumptions in \sect\ref{sect:design-trust}). In the following experiment, we evaluate the robustness of \ourname if this assumption is violated, considering the data poisoning through the injection of attacks during the training process.\\
To do so, we conducted an experiment using the setup \datasetHomeFive and injected different numbers of events that are part of the Light-Flickering attack into the benign training data and evaluated the resulting model. In particular, we considered this attack as it was the attack with the lowest reconstruction scores (i.e., the closest to benign data and so the worst case possible).

\begin{figure}[bt]
	\centering
	\includegraphics[width=0.8\columnwidth]{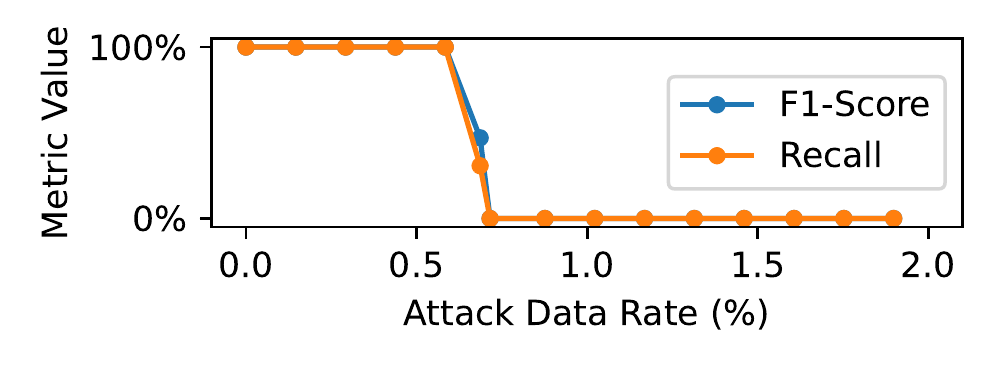}
	\caption{Evaluation of \ourname depending on the amount of poisoned data in the training set of \datasetHomeFive}
	\label{fig:eval-results-poisoning}
\end{figure}

As Fig.~\ref{fig:eval-results-poisoning} shows, \ourname demonstrates robustness characteristics against this kind of data poisoning while the overall poisoned data is lower than 0.6\% of the total number of events. Therefore, \adversary would need to manually perform this attack 42 times, until \ourname cannot detect the Light Flickering anymore.

\section{Computational Setup}
\label{app:computational-setup}
For the evaluation, the events were collected offline. All event traces were evaluated on a server running Debian 10, with 1 TB main memory, 64 physical cores, provided by an AMD EPYC 7742 processor, and 4 NVIDIA Quadro RTX 8000. For the machine learning experiments, we leverage the Scikit-Learn library~\cite{scikitlearn} and Pytorch~\cite{pytorch} for implementing the neural networks.

\end{document}